%% file: Wino_dSphs.tex
\documentclass[12pt]{article}
\usepackage{amsmath, amssymb, slashed, epsf, color, graphicx}

\definecolor{mygreen} {rgb} {0,.7, 0 }

\begin{document}
\begin{titlepage}
\begin{center}

\hfill IPMU14-0121 \\
\hfill ICRR-Report-682-2014-8 \\
\hfill \today

\vspace{1.5cm}
{\large\bf Wino Dark Matter and Future dSph Observations}

\vspace{2.0cm}
{\bf Biplob Bhattacherjee}$^{(a)}$,
{\bf Masahiro Ibe}$^{(a, b)}$,
{\bf Koji Ichikawa}$^{(a)}$, \\
{\bf Shigeki Matsumoto}$^{(a)}$
and
{\bf Kohei Nishiyama}$^{(a)}$

\vspace{1.0cm}
{\it
$^{(a)}${Kavli IPMU (WPI), University of Tokyo, Kashiwa, Chiba 277-8583, Japan} \\
$^{(b)}${ICRR, University of Tokyo, Kashiwa, Chiba 277-8582, Japan}
}

\vspace{2.0cm}
\abstract{
We discuss the indirect detection of the wino dark matter utilizing gamma-ray observations of dwarf spheroidal galaxies (dSphs). After carefully reviewing current limits with particular attention to astrophysical uncertainties, we show prospects of the wino mass limit in future gamma-ray observation by the Fermi-LAT and the GAMMA-400 telescopes. We find that the improvement of the so-called $J$-factor of both the classical and the ultra-faint dSphs will play a crucial role to cover whole mass range of the wino dark matter. For example, with $\delta (\log_{10}J) = 0.1$ for both the classical and the ultra-faint dSphs, whole wino dark matter mass range can be covered by 15\,years and 10\,years data at the Fermi-LAT and GAMMA-400 telescopes, respectively.}

\end{center}
\end{titlepage}
\setcounter{footnote}{0}

\input{Introduction}

\input{Review}

\input{Flux}

\input{Results}

\input{Summary}

\section*{Acknowledgments}

M.I and S.M. were supported by the Grant-in-Aid for Scientific research from the Ministry of Education, Science, Sports, and Culture (MEXT), Japan (No. 2470151 for M.I., and No. 26287039 for M.I. \& S.M.) and from the Japan Society for the Promotion of Science (JSPS), No. 26287039 (M.I.). This work was also supported by the World Premier International Research Center Initiative, MEXT, Japan.

\input{Reference}

\end{document}

%% file: Introduction.tex
\section{Introduction}
\label{sec: intro}

Since the discovery of a new boson at the Large Hadron Collider (LHC)~\cite{Aad:2012tfa}, which seems strongly to be the Higgs boson of the standard model (SM), people have started examining candidates of new physics beyond the SM more closely. One of the most striking hints from the discovery is that its mass is observed at about $126$\,GeV, which indicates that the new physics behind the Higgs mechanism is presumably described by a weakly-interacting theory. Among several weakly interacting extensions of the SM, supersymmetry (SUSY) has been considered so far as the most promising candidate. When SUSY particles exist within a TeV range as expected in the pre-LHC era, however, the Higgs boson mass of $126$\,GeV is difficult to be achieved in the minimal supersymmetric extension of the SM. Rather, larger SUSY breaking effects are mandatory to push up the Higgs boson mass, which in turn requires the typical mass scale of sparticles to be much higher than $1$\,TeV~\cite{Okada:1990vk}. Such high-mass sparticles are actually not only compatible with null-observations of new physics signals at the LHC experiment, but also ameliorate the problem of too large SUSY contributions to flavor-changing neutral current (FCNC) processes.

An apparent downside of high-mass sparticles is the loss of a good candidate for dark matter. When the dark matter is one of the sparticles with the mass much larger than $1$\,TeV, its predicted mass density is too high to be consistent with the observation~\cite{Ade:2013zuv}. This problem is, however, naturally resolved in a class of models of supergravity mediation if the SUSY breaking sector does not include any singlet fields~\cite{Giudice:1998xp, Wells:2004di}. In the models, all scalar particles acquire their masses of the order of the gravitino mass via tree level interactions, while gaugino masses are dominated by one-loop anomaly mediated contributions~\cite{Giudice:1998xp, Randall:1998uk}.\footnote{Origin of the Higgsino mass, the $\mu$-term, is model dependent even in the models. For example, in the pure gravity mediation model~\cite{Ibe:2006de}--\cite{Ibe:2012hu} or the minimal split SUSY model~\cite{ArkaniHamed:2012gw}, the $\mu$-term is generated via a tree level interaction to the $R$-symmetry breaking sector\,\cite{Inoue:1991rk}.} On top of these features, the models predict the lightest supersymmetric particle (LSP) to be the almost pure neutral wino in most parameter space. The neutral wino is known to be a good candidate for a weakly interacting massive particle (WIMP) dark matter when its mass is of ${\cal O}(1)$\,TeV~\cite{Moroi:1999zb}--\cite{Hisano:2006nn}. Therefore, when the gravitino mass is in the range of tens to hundreds TeV range, we can realize a hierarchical spectrum appropriate to explain the observed Higgs boson mass while having a good dark matter candidate. These models are now called high-scale SUSY breaking models.\footnote{For discussions related to the models, see also, for example, the following papers~\cite{Acharya:2007rc}--\cite{Evans:2013nka}.}

Now, the most important question is how and when we will confirm/refute the high-scale SUSY breaking models. Indirect investigations of the heavy scalar sparticles through FCNC processes or electric dipole moments (EDM) of SM particles will play important roles to test some portion of the parameter space~\cite{Hisano:1995cp}, though their signals depend highly on physics behind the flavor/CP structure of squarks and sleptons. Collider experiments such as the LHC and the international linear collider (ILC) will also play some roles as far as gaugino masses are within their accessible ranges~\cite{Tsukamoto:1993gt}--\cite{Aad:2013yna}. However, it is difficult to cover all the parameter region of the model due to unknown model dependences as well as limited energy reaches of the collider experiments. In other words, even if no signals are observed in these experiments, we will not be able to rule out the models completely.

Indirect detections of the wino dark matter, on the other hand, are more hopeful because relevant wino properties are less sensitive to the details of other sparticle masses.\footnote{Direct detection of the wino dark matter is, on the contrary, not hopeful because the wino dark matter scatters off a nucleon at one-loop level. Its typical cross section is estimated to be about 10$^{-47}$cm$^2$~\cite{Hisano:2012wm}, which is beyond the scope of near future experiments~\cite{Aprile:2012nq}.} As a further bonus, the annihilation cross section of the wino dark matter is significantly boosted by the Sommerfeld effect when its mass exceeds $1$\,TeV~\cite{Hisano:2003ec}. Here, let us summarize indirect detections of the wino dark matter. Indirect detections utilizing charged particle fluxes (anti-proton, anti-deuteron, electron, positron, etc.) produced by dark matter annihilations suffer from the large systematic uncertainty of how the charged particles propagate in our galaxy~\cite{Bergstrom:1999jc}. It is thus difficult to completely rule out the models when no signal is observed unless the uncertainty is understood very well. The indirect detection of neutrino flux does not suffer from such an uncertainty in propagation. The acceptance of the neutrino signal is, unfortunately, too low to completely test the models in near future~\cite{Moroi:2011ab}. As a result, the indirect detection utilizing gamma-rays is the best suited for testing the wino dark matter. Gamma-ray signals do not suffer from the uncertainty in propagation and their analysis procedure and techniques are well established.

In this article, we discuss the indirect detection of the wino dark matter utilizing gamma-ray observations of milky-way satellites in particular dwarf spheroidal galaxies (dSphs). Let us comment here that, since gamma-ray travels in a straight line, there are actually several targets which can be used to detect the wino dark matter, such as the central galactic region (CGR) of our galaxy, milky-way satellites, galactic clusters, and the diffused component of gamma-rays. The signal from dark matter annihilation at the CGR, however, suffers from a significant uncertainty due to limited knowledge of dark matter profile and astrophysical background at the region~\cite{Hooper:2012sr}.\footnote{The use of the monochromatic gamma-ray helps to reduce the background gamma-ray and enhances the detectability of the dark matter signal as performed by H.E.S.S. experiment~\cite{Abramowski:2013ax}. With a huge uncertainty of the dark matter profile, however, only a small portion of the wino mass range can be excluded (see references~\cite{Cohen:2013ama,Fan:2013faa} for related discussions).} The signal from galactic clusters is less certain due to unknown boost factors~\cite{Gao:2011rf}, and it seems difficult to test the models in near future~\cite{Ackermann:2010rg}. Observation of diffuse gamma-rays to detect the wino dark matter is almost free from large systematic uncertainties caused by dark matter profile and astrophysical background. However, its signal is weak compared to others~\cite{Abdo:2010dk}. In conclusion, gamma-ray signals from dSphs are the best suited, which are expected to be strong enough to test the wino dark matter, while systematic uncertainties from dark matter profile in each dSph and astrophysical background are much smaller than those of the CGR~\cite{Ackermann:2013yva}. We therefore focus on the indirect detection of the wino dark matter utilizing gamma-ray observations of dSphs in this article, and study whether or not the observations allow us to completely test the high-scale SUSY breaking models in (near) future by carefully investigating astrophysical backgrounds and capabilities of current and future gamma-ray observations (telescopes).

This article is organized as follow. In next section, we review the wino dark matter in the framework of the high-scale SUSY breaking models and summarize present limits on the wino dark matter obtained from the LHC experiment and cosmology. Several topics relevant to the gamma-ray flux from wino dark matter annihilations in dSphs are summarized in section~\ref{sec: flux}, where the annihilation of the wino dark matter, the dark matter density profile inside each dSph, several astrophysical backgrounds against the wino dark matter detection, and the capability of present and future gamma-ray telescopes are carefully discussed. Our final results are shown in section~\ref{sec: results}, where both present and expected future limits on the wino dark matter annihilation obtained from dSph observations are discussed. According to these results, we also consider what kind efforts and/or additional observations are required to completely test the high-scale SUSY breaking models. Section~\ref{sec: summary} is devoted to the summary of our discussions.

%% file: Review.tex
\section{Wino dark matter}
\label{sec: review}

We briefly review the high-scale SUSY breaking models paying special attention to how the wino LSP is realized in a class of models of supergravity mediation without a singlet SUSY breaking field. We also discuss current limits on the wino dark matter obtained by the LHC experiment (which gives the lower limit on the wino mass) and cosmology (which gives the upper limit on the mass).

\subsection{Wino in High-scale SUSY models}

In models with supergravity mediation, scalar bosons generically obtain their soft-SUSY breaking mass terms via tree-level interactions in supergravity. With a generic K\"ahler potential, all the masses are expected to be of the order of the gravitino mass, which is denoted by $m_{3/2}$. Origins of the $\mu$ and the $B$ terms are model dependent, and we assume that they are of the order of the gravitino mass in following discussions. It should be stressed that such $\mu$ and $B$ terms are naturally realized in the pure gravity mediation model~\cite{Ibe:2006de}--\cite{Ibe:2012hu} and the minimal Split SUSY model~\cite{ArkaniHamed:2012gw} even in the absence of any singlet fields in the SUSY breaking sector.

For gaugino masses, on the contrary, tree-level contributions in supergravity are extremely suppressed, since they require a singlet SUSY breaking field. At one-loop level, however, the gaugino masses are generated without such a singlet SUSY breaking field, via anomaly mediated contributions~\cite{Giudice:1998xp, Randall:1998uk}.\footnote{Tri-linear couplings (i.e. $A$-terms) are also suppressed at tree-level in the absence of a singlet SUSY breaking field and they are dominated by anomaly mediated contributions.} In addition, electroweak gauginos also receive other contributions from the heavy Higgsino threshold effect at one-loop level~\cite{Giudice:1998xp, Gherghetta:1999sw}. Putting these one-loop contributions together, the gaugino masses at the energy scale of ${\cal O}(m_{3/2})$ are given by
\begin{eqnarray}
M_1 &=& g_1^2/(16 \pi^2) (33/5) (m_{3/2}+ L/11)\ , \label{eq:M1} \\
M_2 &=& g_2^2/(16 \pi^2) (m_{3/2} + L)\ , \label{eq:M2} \\
M_3 &=& g_3^2/(16 \pi^2) (-3) \, m_{3/2}\ , \label{eq:M3}
\end{eqnarray}
where subscripts `1, 2, 3' correspond to the SM gauge groups $U(1)_Y$, $SU(2)_L$, and $SU(3)_c$ with $g_1$, $g_2$, and $g_3$ being corresponding gauge coupling constants, respectively. Terms proportional to $m_{3/2}$ denote the anomaly mediated contributions, while those proportional to $L$ are the Higgsino threshold contributions,
\begin{eqnarray}
L \equiv \mu \, \sin2\beta
\frac{m_A^2}{(|\mu|^2 - m_A^2)} \ln \frac{|\mu|^2}{m_A^2}\ .
\end{eqnarray}
Here, $m_A$ denotes the mass of the pseudoscalar Higgs boson and $\tan \beta$ is the ratio of the vacuum expectation values between up-type and down-type Higgs doublet fields. As shown in reference~\cite{Ibe:2012hu}, typical values of $\tan\beta$ and $L$ are predicted to be ${\cal O}(1)$ and ${\cal O}(m_{3/2})$, respectively, when $\mu = {\cal O}(m_{3/2})$ and $B={\cal O}(m_{3/2})$. As a result, we immediately find that the gaugino masses are in hundreds GeV to a TeV range when the gravitino mass is fixed to be ${\cal O}$(10--100)\,TeV. This value of the gravitino mass is favored by the observed Higgs boson mass.

By integrating out heavy particles (i.e. sfermions, Higgsinos and heavy Higgs bosons) and taking care of renormalization group running down to the TeV scale, we obtain the low-energy effective lagrangian of the gauginos,
\begin{eqnarray}
{\cal L}_{\rm eff} = {\cal L}_{\rm SM}
+ \bar{\tilde{g}} (i\slashed{D} - m_{\tilde{g}}) \tilde{g}
+ \bar{\tilde{b}} (i\slashed{\partial} - m_{\tilde{b}}) \tilde{b}
+ \bar{\tilde{w}} (i\slashed{D} - m_{\tilde{w}}) \tilde{w}
+ {\cal L}_{\rm H.O.}\ .
\end{eqnarray}
Here, $\tilde{g}$, $\tilde{b}$, and $\tilde{w}$ represent gluino, bino, and wino fields, respectively, with $\slashed{D}$ being their covariant derivatives. The standard model lagrangian is denoted by ${\cal L}_{\rm SM}$. The term ${\cal L}_{\rm H.O.}$ is composed of higher dimensional operators induced by integrating out the heavy fields, which play important roles for gluino and bino decays. Gaugino masses, $m_{\tilde{g}, \, \tilde{b}, \, \tilde{w}}$, obtained by solving renormalization group equations with boundary conditions in equation\,(\ref{eq:M1})--(\ref{eq:M3}) are given by,
\begin{eqnarray}
m_{\tilde{g}} &\simeq& 2.5 \times 10^{-2} \,
m_{3/2} \, (1 - 0.13 \, \delta_{32} - 0.04 \, \delta_{\rm SUSY})\ , \\
m_{\tilde{b}} &\simeq& 9.6 \times 10^{-3} \,
(m_{3/2} + L/11) \, (1 + 0.01 \, \delta_{\rm SUSY})\ , \\
m_{\tilde{w}} &\simeq& 3.0 \times 10^{-3} \,
(m_{3/2} + L) \, (1 - 0.04 \, \delta_{32} + 0.02 \, \delta_{\rm SUSY})\ .
\end{eqnarray}
$\delta_{\rm SUSY} = \ln(M_{\rm SUSY}/100 \,{\rm TeV})$ for all the gauginos, while $\delta_{32} = \ln(m_{3/2}/100 \, {\rm TeV})$ for the gluino and $\delta_{32} = \ln[(m_{3/2} + L) /100 \,{\rm TeV}]$ for the wino~\cite{Ibe:2012hu}.

As can be seen in the above mass formulae, the wino is the LSP unless $L$ is too large compared to $m_{3/2}$. Notice that, if $L$ is too large and the bino is the LSP, such a parameter region has already been phenomenologically excluded, because the bino dark matter would result in too much relic density to be consistent with the observed one.\footnote{When wino and the bino masses are highly degenerated, there is a parameter region consistent with the observation with the bino being dark matter. This region can be probed in another way instead of the one discussed in this article. See reference~\cite{Ibe:2013pua} for more details.}. It is also worth noting that the mixing between the bino and the wino caused by the electroweak symmetry breaking is negligibly small, since the Higgsino mass is ${\cal O}(m_{3/2})$ and much larger than the gaugino masses. We can therefore safely ignore the existence of the gluino and the bino as well as higher dimensional operators involved in ${\cal L}_{\rm eff}$ when physics concerns only the wino dark matter. The effective lagrangian for the dark matter is then simply approximated by
\begin{eqnarray}
{\cal L}_{\rm eff} \simeq
{\cal L}_{\rm SM} + \bar{\tilde{w}} (i\slashed{D} - m_{\tilde{w}}) \tilde{w}\ .
\end{eqnarray}
In this effective theory, there is only one new physics parameter, $m_{\tilde{w}}$. 

\subsection{Constraints on LSP wino}

We first discuss a limit on the wino mass ($m_{\tilde{w}}$) obtained by the LHC experiment. Broadly speaking, there are two possible ways to produce the wino. First one is the pair production of the gluino and its subsequent decay into two quarks and a charged/neutral wino, which leads to a conventional multiple jets plus missing transverse energy signature. This process, however, gives a limit on the gluino mass rather than the wino mass. In other words, if the gluino mass is heavier than $2.3$\,TeV, we do not have any limit on $m_{\tilde{w}}$ even at $14$\,TeV running~\cite{Bhattacherjee:2012ed}. 

A more distinctive possibility comes from the direct wino production through electroweak interactions, namely the Drell-Yan process, $pp \to g/q + \tilde{w}^0 \tilde{w}^\pm \, (\tilde{w}^\pm \tilde{w}^\mp)$. Since the charged wino is highly degenerate with the neutral wino in mass, which is about $170$\,MeV, it decays mainly into a neutral wino and a soft pion (that is hardly detected at the LHC) with a long lifetime. Its decay length (without the Lorentz boost factor) is estimated to be about $7$\,cm, which is almost independent of $m_{\tilde{w}}$. Thus, once the charged wino is produced, it can travel about ${\cal O}(10)$\,cm before it decays, leaving disappearing charged track(s) inside inner detectors. As a result, the process predicts a mono-jet plus missing transverse energy signature accompanied by disappearing charged track(s) caused by the charged wino(s).

From the theoretical side, the mass difference between the charged and the neutral winos has been calculated at the two-loop level and it enables us to predict the signal accurately~\cite{Ibe:2012sx}. From the experimental side, the ATLAS collaboration has already reported an analysis on this process using $20$\,fb$^{-1}$ data at $8$\,TeV running~\cite{Aad:2013yna}. The absence of significant deviation from SM backgrounds puts a limit,
\begin{eqnarray}
m_{\tilde{w}} > 270~{\rm GeV} \qquad {\rm (95\% \, C.L.)}.
\end{eqnarray}
This is a very robust limit because it does not depend on other sparticle masses such as the gluino mass and is applicable as long as the wino is a stable LSP. This analysis is expected to be still available at the $14$\,TeV running. The limit is expected to be extended up to $500$\,GeV with $100$\,fb$^{-1}$ data~\cite{LHCProspect}.

We next consider a limit on the wino mass obtained by cosmology. Here, we consider two possibilities to produce the winos in the early universe. One is the traditional thermal production and the other is the non-thermal production from the late-time decay of gravitinos into winos. The thermal contribution to the wino relic abundance, $\Omega_{\rm TH} h^2$, has been estimated in references~\cite{Hisano:2006nn, Cirelli:2007xd}, where all coannihilation processes as well as the Sommerfeld effect were taken into account. It then turns out that $\Omega_{\rm TH} h^2$ explains the observed abundance when $m_{\tilde{w}}$ is $2.8$--$2.9$\,TeV. The non-thermal contribution to the abundance, $\Omega_{\rm NT} h^2$, on the other hand, depends not only on the wino mass $m_{\tilde{w}}$ but also on the reheating temperature $T_R$ after inflation. When the temperature is higher, the more the gravitino is produced, and hence, the contribution is larger. $\Omega_{\rm NT} h^2$ is quantitatively estimated to be $\Omega_{\rm NT} h^2 \simeq 0.16 \, (m_{\tilde{w}}/300 \, {\rm GeV}) \, (T_R/10^{10} \, {\rm GeV})$~\cite{Moroi:1999zb, Gherghetta:1999sw}.

As a minimal setup, we assume that the wino dark matter produced either thermally or non-thermally by the gravitino decay explains the observed dark matter density, i.e. $\Omega_{\rm TH} h^2 + \Omega_{\rm NT} h^2 = \Omega_{\rm DM}^{\rm (obs.)} h^2$, and assume no entropy production in the universe at the later time. For $m_{\tilde{w}} \simeq 2.8$--$2.9$\,TeV, the reheating temperature is required to be low so that the thermally produced wino dominates the dark matter density. For a lighter wino, on the other hand, the non-thermally produced wino dominates the dark matter density by setting $T_R$ appropriately. From these arguments, we immediately find an upper limit on $m_{\tilde{w}}$,
\begin{eqnarray}
m_{\tilde{w}} < 2.9~{\rm TeV} \ , \qquad {\rm (95\% \, C.L.)}\ ,
\label{eq: upper limit on wino mass}
\end{eqnarray}
where we imposed the latest result on $ \Omega_{\rm DM}^{\rm (obs.)} h^2$~\cite{Ade:2013zuv}. The limit is again very robust because it can be applied as long as the wino is the stable LSP. 

It is also worth pointing out that the upper limit becomes stronger down to $m_{\tilde{w}} \lesssim 1~{\rm TeV}$ when we impose $T_R$ to be higher than about $2 \times 10^9$\,GeV as required by the traditional scenario of thermal leptogenesis~\cite{Fukugita:1986hr}, although much lower reheating temperature is allowed in more generic leptogenesis scenarios. Incidentally, as emphasized in reference~\cite{Harigaya:2014pqa}, the observed tensor fraction in the cosmic microwave background radiation by the BICEP2 collaboration~\cite{Ade:2014xna} supports the reheating temperature consistent with thermal leptogenesis. In fact, the observed tensor fraction points to the inflaton mass to be ${\cal O}(10^{13})$\,GeV when it is interpreted in the simplest chaotic inflation model with a quadratic potential\,\cite{Linde:1983gd}.\footnote{For a construction of the chaotic inflation in supergravity, see references~\cite{Kawasaki:2000yn, Kallosh:2011qk}.} In this case, the reheating temperature $T_R$ is predicted to be around the favorable reheating temperature, i.e. $10^{9-10}$\,GeV, when the inflaton decays into radiation via dimension five operators suppressed by the Planck scale.

There are actually many other experiments which are able to put limits on the wino mass. Except for indirect detections of the wino dark matter utilizing gamma-rays, however, those limits are weaker or less robust than the ones discussed in this section.\footnote{The indirect detection utilizing cosmic-ray anti-protons is potentially important, as clearly pointed out in reference~\cite{Hryczuk:2014hpa}, when systematic errors associated with the use of the diffusion equation are accurately evaluated. The limit on $m_{\tilde{w}}$ could be as strong as $m_{\tilde{w}} > 500$\,GeV.} As we will show, the most stringent and robust limit from the indirect detections comes from dSph observations. In following sections, we will therefore discuss the present limit on $m_{\tilde{w}}$ from the dSph observations and consider its prospect on how widely the wino mass region can be probed in (near) future.

%% file: Flux.tex
\section{Gamma-ray flux from wino annihilations}
\label{sec: flux}

As already mentioned in introduction, we focus on the indirect detection of the wino dark matter utilizing gamma-rays from dwarf spheroidal galaxies (dSphs). There are several advantages to consider the dSphs as the target to detect the wino dark matter. First, the measured values of mass-to-light ratio of the dSphs are very high and it indicates that they are dark matter rich objects. Second, baryonic gas densities inside the dSphs are very low which reduce astrophysical gamma-ray backgrounds to small values. Finally, they are also relatively nearby from us and the measurement of velocity dispersions inside each dSph allows us to estimate its dark matter profile precisely. The indirect detection of dark matter by the dSph observations is actually known to give a strong and robust limit on various dark matter candidates. In fact, the observations of gamma-rays from the dSphs have already put strong limits on some mass range of the wino dark matter.

At a given energy $E$ of the gamma-ray, the differential gamma-ray flux from wino dark matter annihilations in each dSph in a solid angle $\Delta \Omega$ is given by
\begin{eqnarray}
\Phi (E, \Delta \Omega)
=
\left[
\frac{\langle \sigma v \rangle}{8 \pi m_{\tilde{w}}^2} 
\sum_{f}
{\rm Br}(\tilde{w}^0 \tilde{w}^0 \to f)
\left(\frac{dN_\gamma}{dE}\right)_f
\right]
\left[
\rule{0ex}{4ex}
\int_{\Delta \Omega} d\Omega \int_{l.o.s.} dl \,
\rho^2(l, \Omega)
\right].
\label{eq: flux formula}
\end{eqnarray}
With $v$ being the relative velocity between incident wino dark matter particles, $\langle \sigma v \rangle$ denotes the velocity-averaged total annihilation cross section multiplied by $v$, which can be well approximated by the value in the vanishing velocity limit as long as $m_{\tilde \omega}\,v \ll m_{Z,W}$.\footnote{There is another region in which we cannot neglect the velocity dependence on $\sigma v$; the region where the binding energy of the wino bound state is almost zero. Since this region has already been excluded due to the huge annihilation cross, we do not discuss it any more.} ${\rm Br}(\tilde{w}^0 \tilde{w}^0 \to f)$ denotes the branching fraction of the annihilation into a final-state $f$, and $(dN_\gamma/dE)_f$ is the differential number density of photons for a given final state $f$, (i.e. the fragmentation function). The dark matter profile inside dSph is denoted by $\rho(l, \Omega)$. The part in first parenthesis is determined only by particle physics, while the second one, which is called \underline{the $J$-factor $J(\Delta \Omega)$}, is from astrophysics. We will discuss below both factors separately.

\subsection{Particle physics factor}

\begin{figure}[t]
\begin{center}
\includegraphics[scale=0.36]{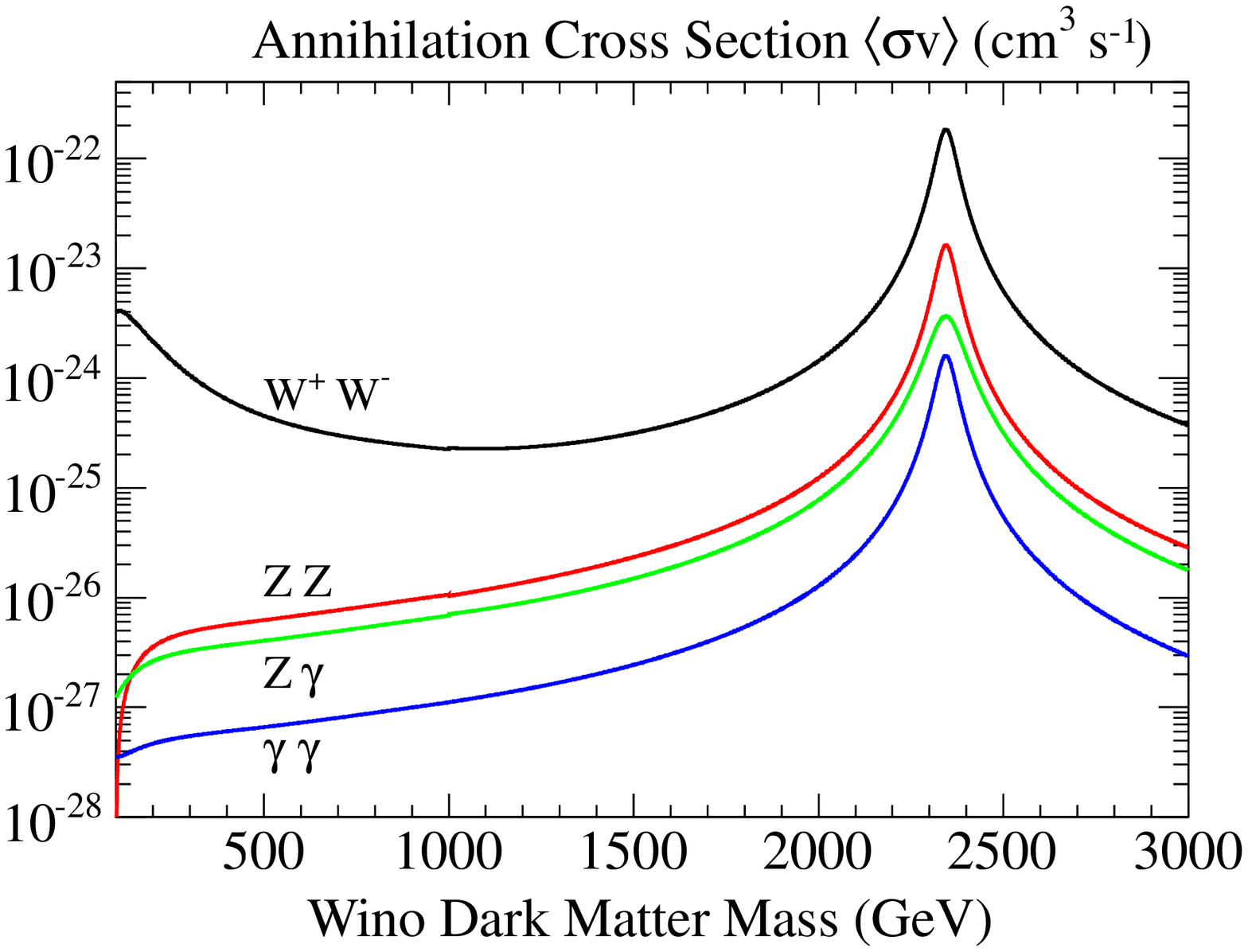}
~~
\includegraphics[scale=0.36]{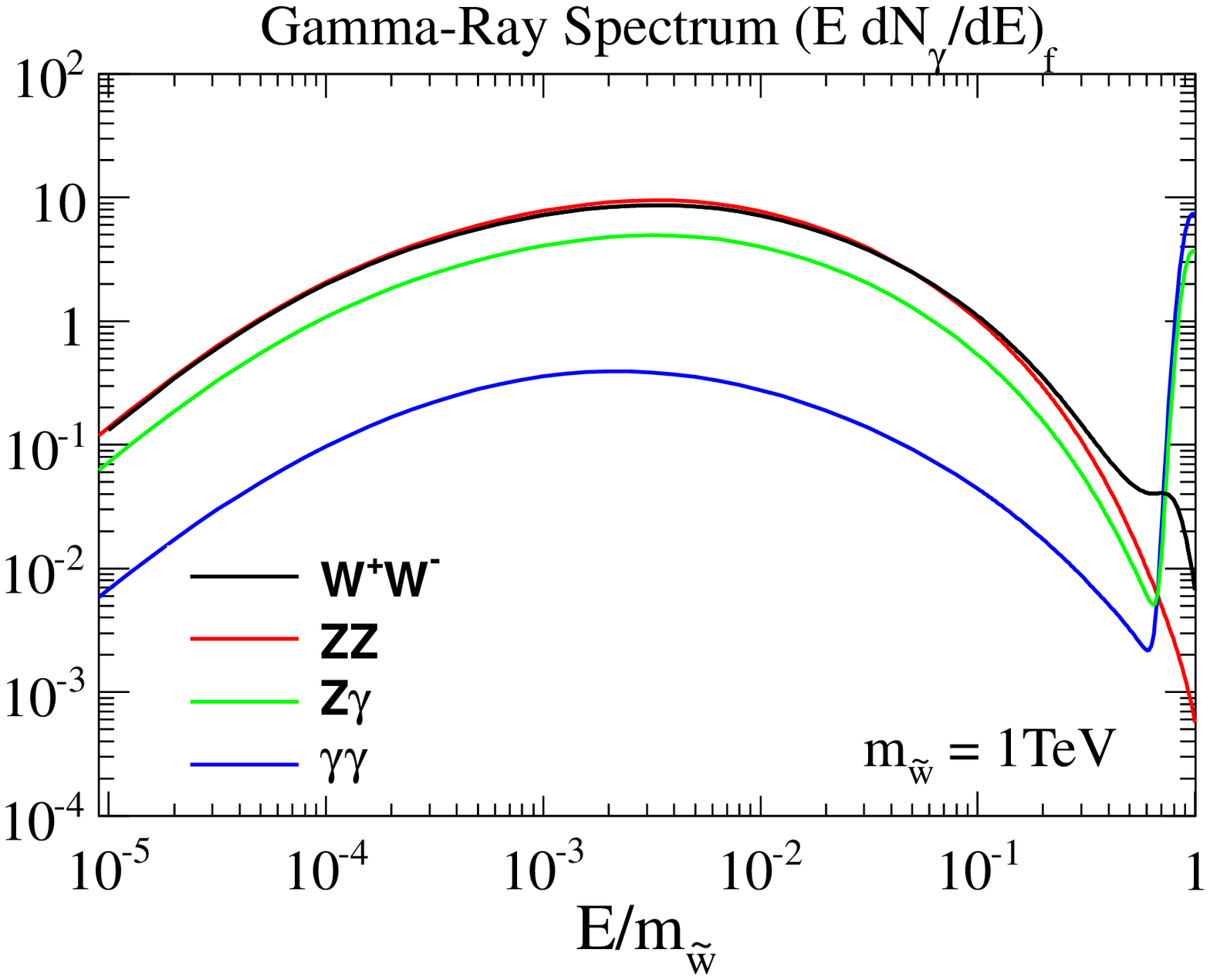}
\caption{\small\sl
{\bf Left panel:} Annihilation cross sections of the wino dark matter for processes $\tilde{w}^0 \tilde{w}^0 \to WW$, $ZZ$, $Z\gamma$, and $\gamma\gamma$ as a function of the wino mass~\cite{Hryczuk:2011vi}. 
{\bf Right panel:} Normalized fragmentation functions, $E (dN_\gamma/dE)_f$, for final states $f = WW$, $ZZ$, $Z\gamma$, and $\gamma\gamma$ as a function of $E$ in unit of $m_{\tilde{w}}$. Gaussian smearing with $\Delta E/E = 0.1$ was applied.}
\label{fig: wino annihilation}
\end{center}
\end{figure}

The wino dark matter dominantly self-annihilates into $W$ boson pair ($WW$) through the process with the $t$-channel exchange of the charged wino. The dark matter can also annihilate into $Z$ boson pair ($ZZ$), $Z$ boson plus photon ($Z\gamma$), and two photons ($\gamma \gamma$) through one-loop processes. Annihilations into fermion pairs are, on the contrary, suppressed due to angular momentum and CP conservations. When the wino dark matter is much heavier than the weak gauge bosons, exchanging the bosons between incident wino dark matter particles causes long-range forces, which leads to the modification of incident wave functions from the plane-waves. The annihilation cross section is as a result enhanced by a few orders of magnitude compared to the leading one, which is called the Sommerfeld effect~\cite{Hisano:2003ec}. Calculation of the annihilation cross section is then divided into two parts: One is the calculation of annihilation amplitudes, which is the same as the one obtained in usual perturbation theory (short-distant part). The other is the calculation of enhancement factors caused by the Sommerfeld effect, which is almost one for low mass region while much larger than one for high mass region (long-distant part). At present, the short-distant and the long-distant parts have been calculated to the next leading and to the leading orders, respectively, in each annihilation mode~\cite{Hryczuk:2011vi}. Resultant annihilation cross sections are shown in the left panel of Fig.~\ref{fig: wino annihilation}.

Weak gauge bosons from the wino dark matter annihilation first decay into quarks, charged leptons, and neutrinos. Once quarks are produced, they are fragmented into various hadrons, eventually producing stable particles such as protons, anti-protons, electrons, positrons, photons, and neutrinos. Photons therefore come as direct annihilation products via processes $Z \gamma$ and $\gamma \gamma$, or as secondary decay products of hadrons (mainly from $\pi^0$ decays). We have as a result monochromatic gamma-rays in the direct annihilations, whereas continuous gamma-rays in the second case. The energy distribution of photons from quark fragmentations is usually calculated by a simulation code such as Pythia~\cite{Sjostrand:2007gs}, which has been developed mainly for collider physics. The code generates the distribution including the effect of QED and QCD final-state radiations. It has been, however, pointed out in reference~\cite{Ciafaloni:2010ti} that the final-state radiations of weak gauge bosons also give sizable contributions to the distribution. The radiations are actually logarithmically enhanced in their soft and collinear parts, and modify the distribution by a factor of two to ten at the photon energy of ${\cal O}$(0.1--10)\,GeV. To incorporate such effects, we have thus used the distribution given in reference~\cite{Cirelli:2010xx}, which includes not only the effect of QED and QCD final-state radiations but also the above electroweak correction at the leading order.\footnote{There are other contributions to the fragmentation function from the Sommerfeld effect and the virtual internal bremsstrahlung~\cite{Bringmann:2007nk}. We have checked that the contributions do not alter our results (less than 5\% level), except the region where the annihilation cross sections peak. The peaked region has already been ruled out by several indirect detection experiments.} The energy-distribution of photons produced in each dark matter annihilation is summarized in so-called the fragmentation function $(dN_\gamma/dE)_f$, which is shown in the right panel of Fig.~\ref{fig: wino annihilation}. In order to sketch a realistic photon spectrum, we have applied a Gaussian smearing to the function with the width of $\Delta E/E = 10\%$ (the typical energy resolution of gamma-ray telescopes). 

Since the annihilation cross sections of the wino dark matter has been calculated with the precision of a few percent level~\cite{Hryczuk:2011vi}, the most dominant systematic error on the particle physics factor in equation~(\ref{eq: flux formula}) comes from the fragmentation functions. In particular, numerical simulations for quark fragmentations tend to give a large error, as discussed in reference~\cite{Cirelli:2010xx, Cembranos:2013cfa}. Fortunately, the wino dark matter annihilates mainly into electroweak gauge bosons, so that the simulations have been tuned very well by collider physics data of e.g. the LEP experiment. The systematic error associated with the particle physics factor is estimated to be at most 10\%, and as a result enough smaller than that of the astrophysical factor. In this article, we therefore use the annihilation cross sections in Fig.~\ref{fig: wino annihilation} and the fragmentation functions provided in reference~\cite{Cirelli:2010xx} assuming no systematic error.

\subsection{Astrophysical factor}
\label{subsec: A factor}

The second term in the flux formula (\ref{eq: flux formula}) is often called the astrophysical $J$-factor, and it is determined by the dark matter density profile inside a dSph, $\rho(l, \Omega)$. Here, $l$ denotes the distance along the line-of-sight and $\Omega$ is the solid angle of an observational cone pointing to the center of the dSph. With $\alpha$ being the angle between the line of sight and the direction to the dSph center, the $J$ factor is defined by
\begin{eqnarray}
J(\alpha) \equiv
\int_{\Delta \Omega(\alpha)} \, d\Omega \int_{l.o.s.} dl \,
\rho^2(l, \Omega),
\label{eq: J-factor}
\end{eqnarray}
where the solid angle is given by $\Delta \Omega(\alpha) = 2 \pi \, (1 - \cos\,\alpha)$.

The dark matter profile is usually evaluated by comparing the mass-model of dSphs and the stellar kinematic data of the dSphs (e.g. velocity dispersions of stellar objects). Since dSphs are dark-matter-rich astrophysical objects, the stellar kinematics are governed mostly by how the dark matter is distributed inside the dSphs, namely the dark matter profile. The profile is generally assumed to be spherically distributed and described by the function~\cite{Hernquist:1990be}:
\begin{equation}
\rho(r)= \rho_s \, (r/r_s)^{-\gamma} \,
[ 1 + (r/r_s)^\alpha ]^{(\gamma - \beta)/\alpha},
\label{eq: DM profile}
\end{equation}
where $r$ is the distance from the dSph center. Parameter $\gamma$ determines the inner slope of the profile (say, cuspy or cored), $\beta$ describes the outer slope, $\alpha$ controls the sharpness of transition from the inner to the outer slopes at a characteristic scale $r_s$, and $\rho_s$ is a normalization factor. The profile is thus completely specified by evaluating five parameters, $\alpha$, $\beta$, $\gamma$, $r_s$ and $\rho_s$, from stellar kinematics. Notice that the so-called NFW profile ($\alpha = 1, \, \beta = 3, \, \gamma = 1$)~\cite{Navarro:1996gj} is adopted in many articles to reduce the free parameters. On the other hand, recent observations suggest another profile possibility which is cored at the center~\cite{Walker:2011zu}. The most typical one is called the Burkert profile ($\alpha \simeq 1.5, \, \beta = 3, \, \gamma = 0$)~\cite{Burkert:1995yz}.

The size of the observational cone in the $J$-factor~(\ref{eq: J-factor}) is usually taken to be around the half-light radius $r_e$ which turns out to be similar to $r_s$, corresponding to the angle $\alpha_e \simeq r_e/d$ with $d$ being the distance between dSph and us~\cite{Charbonnier:2011ft}. This choice minimizes the systematic error on the factor. That is, the choice of a much smaller $\alpha$ than $\alpha_e$ not only reduces the signal flux but also enhances the error of $J$-factor due to the decreasing stellar kinematic data. The much larger $\alpha$ than $\alpha_e$ also enhances the systematic error of the $J$-factor because the effect of dark matter substructures around the dSphs (e.g. dark matter clumps) is expected to contribute to the profile~\cite{Bullock:1999he}. Besides, since the $J$-factor is proportional to dark matter density squared, the factor is not enhanced even if we take larger $\alpha$. According to analysis by the Fermi-LAT collaboration, we take $\alpha = 0.5^\circ$ for all the dSphs in our analysis, which satisfies the above condition.

The eight dSphs, Ursa Minor, Sculptor, Draco, Sextans, Carina, Fornax, Leo\,I, and Leo\,I\hspace{-.1em}I have been discovered before the Sloan Digital Sky Survey (SDSS) observation~\cite{York:2000gk}, and are now called the classical dSphs. In our analysis, we use the four classical dSphs, Ursa Minor, Draco, Sculptor, and Sextans, because their locations are close to us (within 100\,kpc) and give sizable contributions for the wino dark matter search. The $J$-factors of the other classical dSphs are negligibly small for the purpose. Information about the four dSphs is shown in the table below. The median values and the errors of the $J$-factors were obtained by the Bayesian analysis assuming the NFW profile~\cite{Martinez:2013els}, where prior dependence on the posterior probability of the $J$-factors turns out to almost vanish~\cite{Martinez:2009jh}. It is worth emphasizing that, because stellar kinematic data of the four dSphs have already been accumulated enough~\cite{Wolf:2009tu}, the maximum likelihood analysis can also evaluate their $J$-factors well even if we use the most generalized dark matter profile~(\ref{eq: DM profile})~\cite{Charbonnier:2011ft}. The result of the maximum likelihood analysis turns out to be consistent with that from the Bayesian analysis, so that the estimation of the $J$-factors given in the table is robust.

\vspace{0.2cm}
\begin{center}{\footnotesize
\begin{tabular}{lccccc}
& long.\,(deg.) & lat.\,(deg.) & dist.\,(kpc) & $\alpha_s$\,(deg.) &
$\log_{10}[J(0.5^\circ)/({\rm GeV}^2 {\rm cm}^{-5} {\rm sr})]$ \\
\hline
Draco & 86.4 & 34.7 & 76 & $0.25^{+0.15}_{-0.09}$ & $18.8\pm0.16$ \\
Ursa Min. & 105.0 & 44.8 & 76 & $0.32^{+0.18}_{-0.12}$ & $18.8\pm0.19$ \\
Sculptor & 287.5 & -83.2 & 86 & $0.25^{+0.25}_{-0.13}$ & $18.6\pm0.18$ \\
Sextans & 243.5 & 42.3 & 86 & $0.13^{+0.07}_{-0.05}$ & $18.4\pm0.27$ \\
\hline
\end{tabular}
}\\ \end{center}
\vspace{0.4cm}

After the SDSS observation, many fainter dwarf spheroidal galaxies called the ultra-faint dSphs have been discovered. Some of them are located within 10--50\,kpc from us, and their mass-to-light ratios are about ten times larger than those of the classical dSphs. The ultra-faint dSphs are therefore expected to have large $J$-factors, and improve sensitivity of dark matter detection. Their dark matter profiles are, however, not fixed well due to limited stellar kinematic data: only 10--100 stellar kinematic data have been obtained at present for each ultra-faint dSph. The data are too limited to evaluate the dark matter profile by the maximum likelihood analysis, while large prior dependence on the prior probability remains in the Bayesian analysis~\cite{Martinez:2009jh}. Currently, the $J$-factors of the ultra-faint dSphs are estimated by a two-level Bayesian hierarchical model~\cite{Martinez:2013els} in order to avoid arbitrary choice of the prior probability. In this model, all dSphs are assumed to have some common relations among luminosity, maximum circular velocity, and the radius of the velocity, and they are used in the bottom-level prior probability.\footnote{Explicit forms of the relations are determined based on simulations/observations~\cite{Springel:2008cc} and they involve some free parameters. These parameters are evaluated using data of all dSphs.}

Validity of the use of the relations is, however, not guaranteed, because origins of the ultra-faint dSphs are still under debate and it is not clear whether or not such relations hold for all the dSphs. Thus, at this point, the dark matter constraint obtained from the ultra-faint dSphs seems less conservative. On the other hand, many efforts are now being paid to obtain more kinematic data of the ultra-faint dSphs by deeply observing them, and the dSphs will eventually play important roles for detecting dark matter signals in (near) future. We will therefore involve ultra-faint dSphs in our analysis of future prospects, with errors on their $J$-factors being free parameters. Mean values of the $J$-factors are chosen according to results obtained by the two-level Bayesian hierarchical model as a reference (see next section for more details). As is the same reason for the classical dSphs, the following four ultra-faint dSphs are used in our analysis: Segue\,1, Ursa Major\,II, Willman\,1, Coma Berenices, and information about the dSphs are shown below in the form of a table. Note that the mean values and the errors of the $J$-factors (shown by the italic font) are the ones obtained by the two-level Bayesian hierarchical model.

\vspace{0.2cm}
\begin{center}{\footnotesize
\begin{tabular}{lccccc}
& long.\,(deg.) & lat.\,(deg.) & dist.\,(kpc) & $\alpha_{s}$\,(deg.) &
$\log_{10}[J(0.5^\circ)/({\rm GeV}^2 {\rm cm}^{-5} {\rm sr})]$ \\
\hline
Segue\,1 & 220.5 & 50.4 & 23 & $0.40^{+0.86}_{-0.27}$ & ${\it 19.5\pm0.29}$ \\
Ursa Maj.\,II & 152.5 & 37.4 & 32 & $0.32^{+0.48}_{-0.19}$ & ${\it 19.3\pm0.28}$ \\
Willman\,1 & 158.6 & 56.8 & 38 & $0.25^{+0.54}_{-0.17}$ & ${\it 19.1\pm0.31}$ \\
Coma B. & 241.9 & 83.6 & 44 & $0.25^{+0.54}_{-0.17}$ & ${\it 19.0\pm0.25}$ \\
\hline
\end{tabular}
}\end{center}

\subsection{Backgrounds}

We discuss here astrophysical backgrounds against the dark matter signal from various dSphs, which originate in galactic diffuse emissions, isotropic diffuse emissions, and point source emissions. The galactic diffuse emissions come from the decay of neutral pions produced by the collision between the cosmic-ray (CR) and the interstellar medium (ISM), the bremsstrahlung of CR electrons in the ISM, and the inverse Compton scattering off the interstellar radiation field (ISFR). Gamma-ray emissions from large scale structures such as the hard-spectrum lobes (Fermi Bubbles)~\cite{Su:2010qj} and the giant radio loop (Loop I)~\cite{LoopI} also contribute to the diffuse component. Any dSphs we are considering, which are shown in the tables in previous subsection, are not located on the directions of these structures. The isotropic emissions are, on the other hand, composed of several extragalactic contributions: active galactic nucleus, starburst galaxies, gamma-ray bursts, and other unknown sources.\footnote{Dark matter annihilations in our galactic halo and those of extra-galaxies also contribute to the isotropic emissions, though they are smaller than other contributions~\cite{Abdo:2010dk}. Furthermore, since the isotropic emissions are evaluated with the direct use of Fermi-LAT diffuse gamma-ray data as mentioned in following discussion, the contributions do not affect our analysis at all.} Point source emissions are mainly from active galaxies, mostly blazers. Supernova remnants and pulsars are also a part of the contribution. Furthermore, there are a large number of point sources which are not identified yet. 

In our analysis, the astrophysical backgrounds are evaluated based on the background model provided by the Fermi-LAT collaboration. The galactic diffused emissions are estimated by the GALPROP code~\cite{GALPROP} using interstellar gas distributions (mainly HI and HII gasses) for the neutral pion production, the ISRF model for the inverse Compton scattering~\cite{Moskalenko:1998gw}, and the models of the large structures~\cite{Su:2010qj}. The isotropic emissions are, on the other hand, evaluated directly from the observational gamma-ray data of all-sky except the region $|b| < 30^{\circ}$, where $b$ is the galactic latitude: the emissions are obtained by subtracting the galactic diffuse emissions and the point source emissions~\cite{Abdo:2010nz} from the data using the profile likelihood analysis with their normalizations being free parameters. It is worth emphasizing that the Fermi-LAT collaboration estimated the uncertainty of the model by examining the different choice of the magnetic diffusion zone, the ISRF model, and the sky region (to investigate contamination from CR and unrated sources). It then turned out that the choice makes only a small difference, which is in fact smaller than the uncertainty to determine the normalizations in the profile likelihood analysis. 

\begin{figure}[t]
\begin{center}
\includegraphics[scale=0.367]{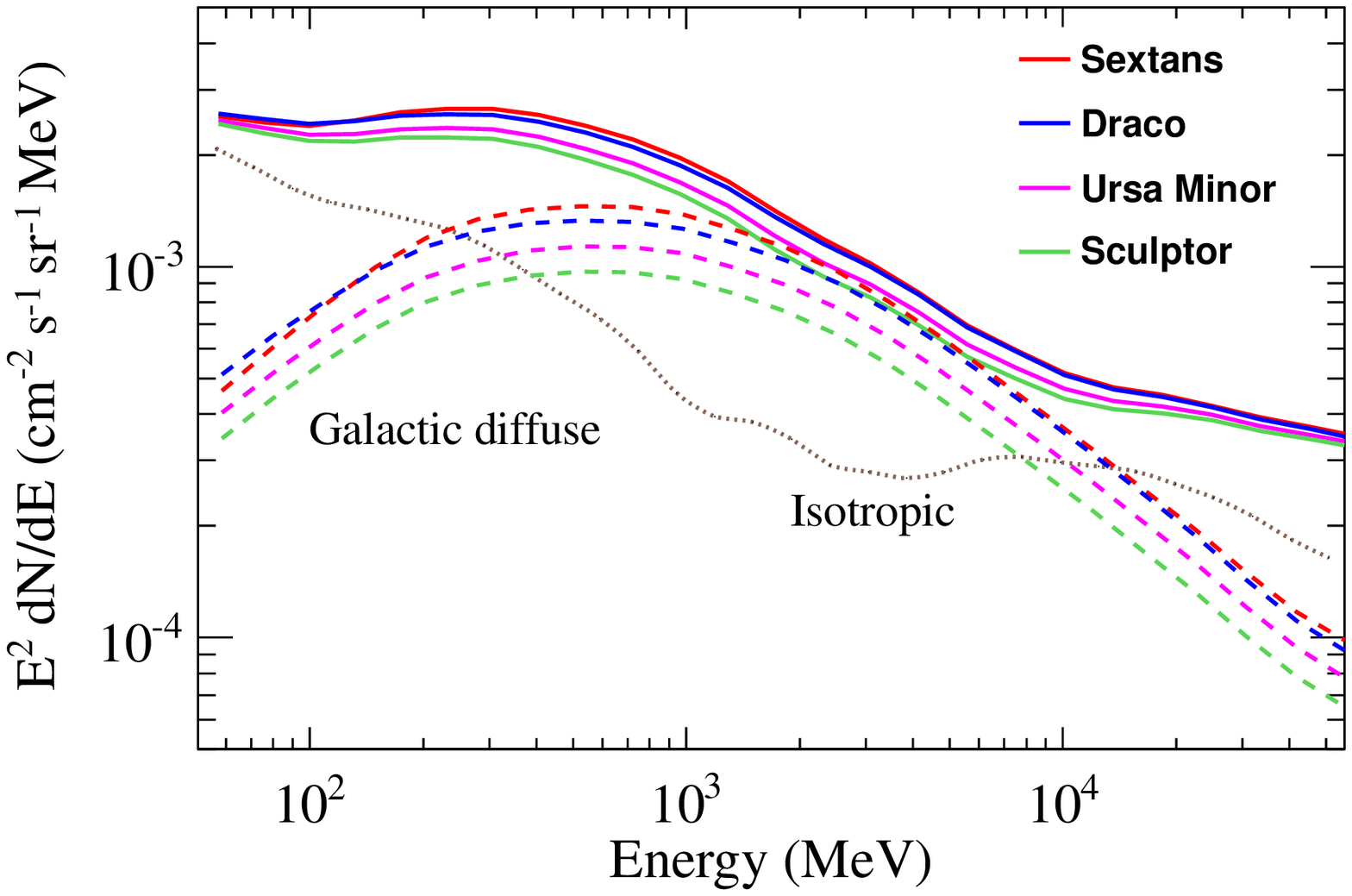}
~
\includegraphics[scale=0.367]{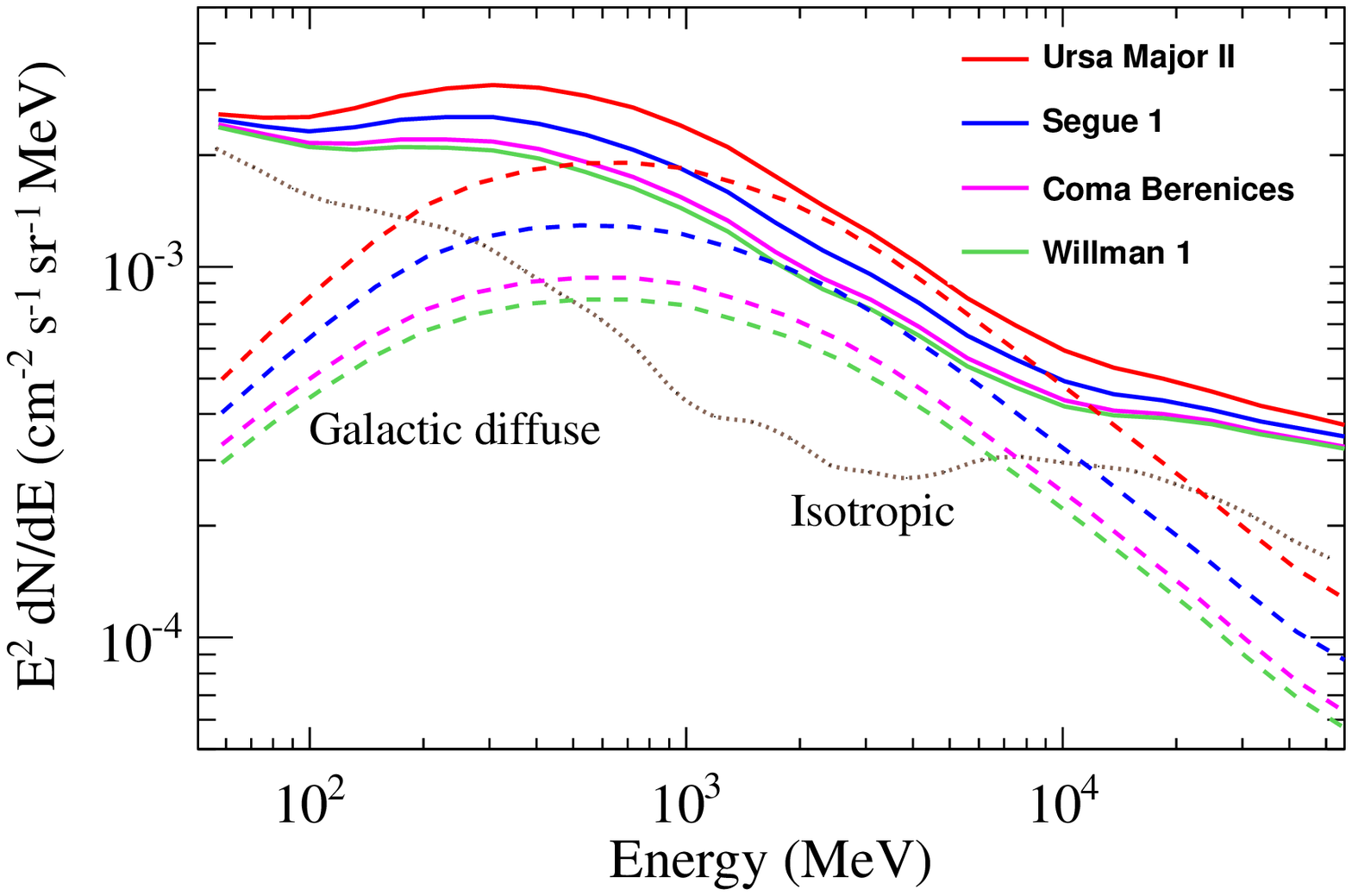}
\caption{\small\sl
Astrophysical background fluxes per unit solid angle averaged over the signal region for classical ({\bf left panel}) and ultra-faint ({\bf right panel}) dSphs as a function of the gamma-ray energy. Galactic and isotropic diffuse components are shown as broken and dotted lines, respectively, while their sums are shown as solid lines.}
\label{fig: backgrounds}
\end{center}
\end{figure}

Among various data of the astrophysical backgrounds provided by the Fermi-LAT collaboration~\cite{FermiBG}, we use the data `gll\_iem\_v05.fit' for the galactic diffuse background, which is obtained based on the highly sophisticated data-classification called `Pass\,7'~\cite{Ackermann:2012kna, Bregeon:2013qba}. The isotropic diffuse background model is, on the other hand, released based on two different selection criteria; `Pass\,7 SOURCE' and `Pass\,7 CLEAN'~\cite{FermiBG}. Though the SOURCE class data gives a larger number of statistics, it contains a significant amount of misidentified CR contributions, especially at the energy region above 1\,GeV~\cite{Ackermann:2012kna}. In order to avoid such a contamination, we take the CLEAN class data `iso\_clean\_v05.txt'. Emissions from point and unrated sources which are overlapped with the signal region (which is defined as a $1.0^\circ \times 1.0^\circ$ square pointing to a dSph in this article) may change the background normalization. Since the change is estimated to be at most ${\cal O}(10)$\% level~\cite{Cholis:2012am}, we neglect their contributions to the astrophysical backgrounds. The background flux is then estimated by integrating galactic and isotropic diffuse emissions over the signal region. The background flux per unit solid angle averaged over the signal region is shown in Fig.~\ref{fig: backgrounds} as a function of the gamma-ray energy for each classical/ultra-faint dSph.\footnote{Since no significant gamma-ray excesses have been observed yet for all the directions of the dSphs, it is good enough to estimate the diffused background averaged over the signal region.} We have checked that the background flux is not altered even if we use larger region: e.g. averaging over a $5^{\circ} \times 5^{\circ}$ square gives at most 10 percent deviation. 

\subsection{Detector capabilities}

\begin{figure}[t]
\begin{center}
\includegraphics[scale=0.75]{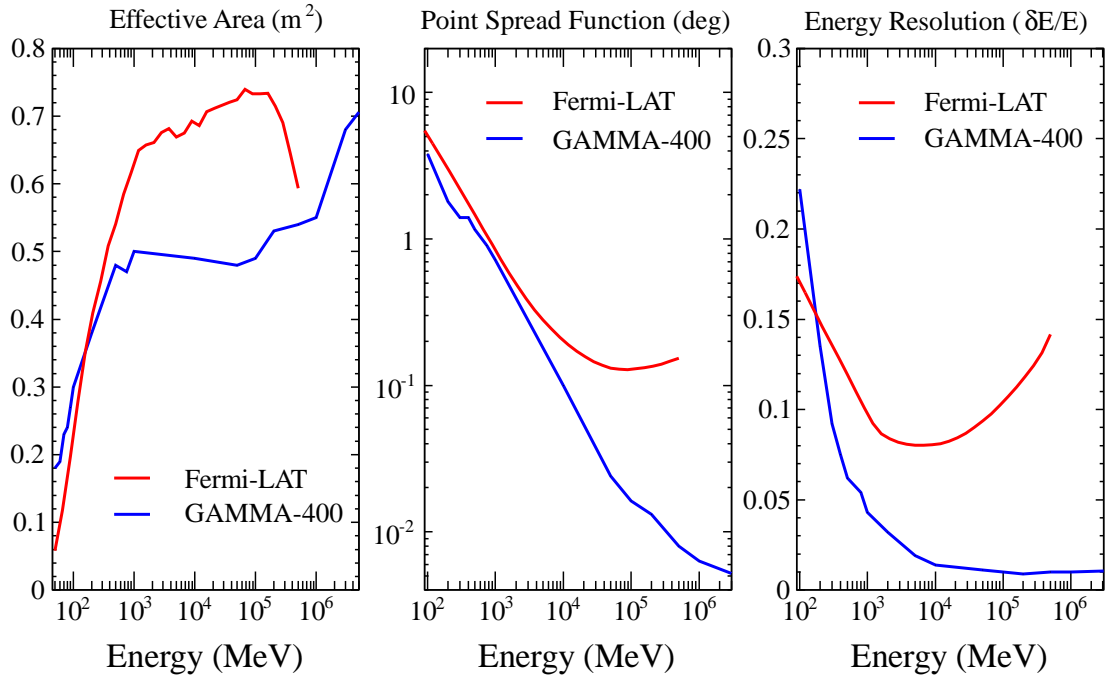}
\caption{\small\sl The effective area {\bf (left panel)}, the point spread function (PSF) {\bf (middle panel)}, and the energy resolution {\bf (right panel)} are shown as a function of the gamma-ray energy (in unit of MeV). Both cases for the Fermi-LAT~\cite{IRFF} and the GAMMA-400 telescopes~\cite{IRFG} are shown in each figure as red and blue lines, respectively.}
\label{fig: IRF}
\end{center}
\end{figure}

The number of signal and background events in actual observations depends on not only their fluxes but also the capability of detectors (gamma-ray telescopes). In our analysis, we consider the Fermi-LAT~\cite{Atwood:2009ez} and the future projected GAMMA-400~\cite{Galper:2012fp} telescopes. Such kind of satellite-borne gamma-ray telescopes can cover the whole sky region and thus efficiently accumulate the signal events from various dSphs.\footnote{The proton rejection factor is also better than other kinds of telescopes, which is estimated to be $10^4$ for the Fermi-LAT telescope and $10^6$ for the GAMMA-400 telescope, respectively.} The most important aspect of the capability is from the effective area, the point spread function (PSF), and the energy resolution; those are often called the instrumental response functions (IRFs). The effective area is determined by the gamma-ray conversion rate induced by a thin foil in the detectors, and depends also on event identification algorithm. The left panel of Fig.~\ref{fig: IRF} is showing the energy dependence of the effective area using the CLEAN class IRF, `P7REP\_CLEAN\_V15'~\cite{IRFF}, for the Fermi-LAT and the IRF in reference~\cite{IRFG} for the GAMMA-400. It can be seen that the area of the GAMMA-400 is 40--100\% smaller than that of the Fermi-LAT, though the energy range covered by the former telescope (0.1--3000\,GeV) is much larger than that of the latter one (0.1--500\,GeV). The PSF is determined by the strip geometry of the detectors and the track reconstruction uncertainty from the multiple scattering of created electrons. The latter factor becomes significant for low-energy gamma-rays, as shown in the middle panel of Fig.~\ref{fig: IRF}. The PSF of the GAMMA-400 is substantially smaller than that of the Fermi-LAT when $E \gtrsim 10^4$\,MeV. The energy resolution is determined by the energy loss inside the tracker and the shower leakage inside the electromagnetic calorimeter. Though very energetic gamma-rays with $E \gtrsim 100$\,GeV rarely deposit their energies inside the calorimeter, the energies are deduced from a sophisticated shower imaging analysis. The right panel of Fig.~\ref{fig: IRF} is showing the energy resolution. It can be seen that the GAMMA-400 covers wider energy range and gives better resolution thanks to the thick calorimeter.

\begin{figure}[t]
\begin{center}
\includegraphics[scale=0.367]{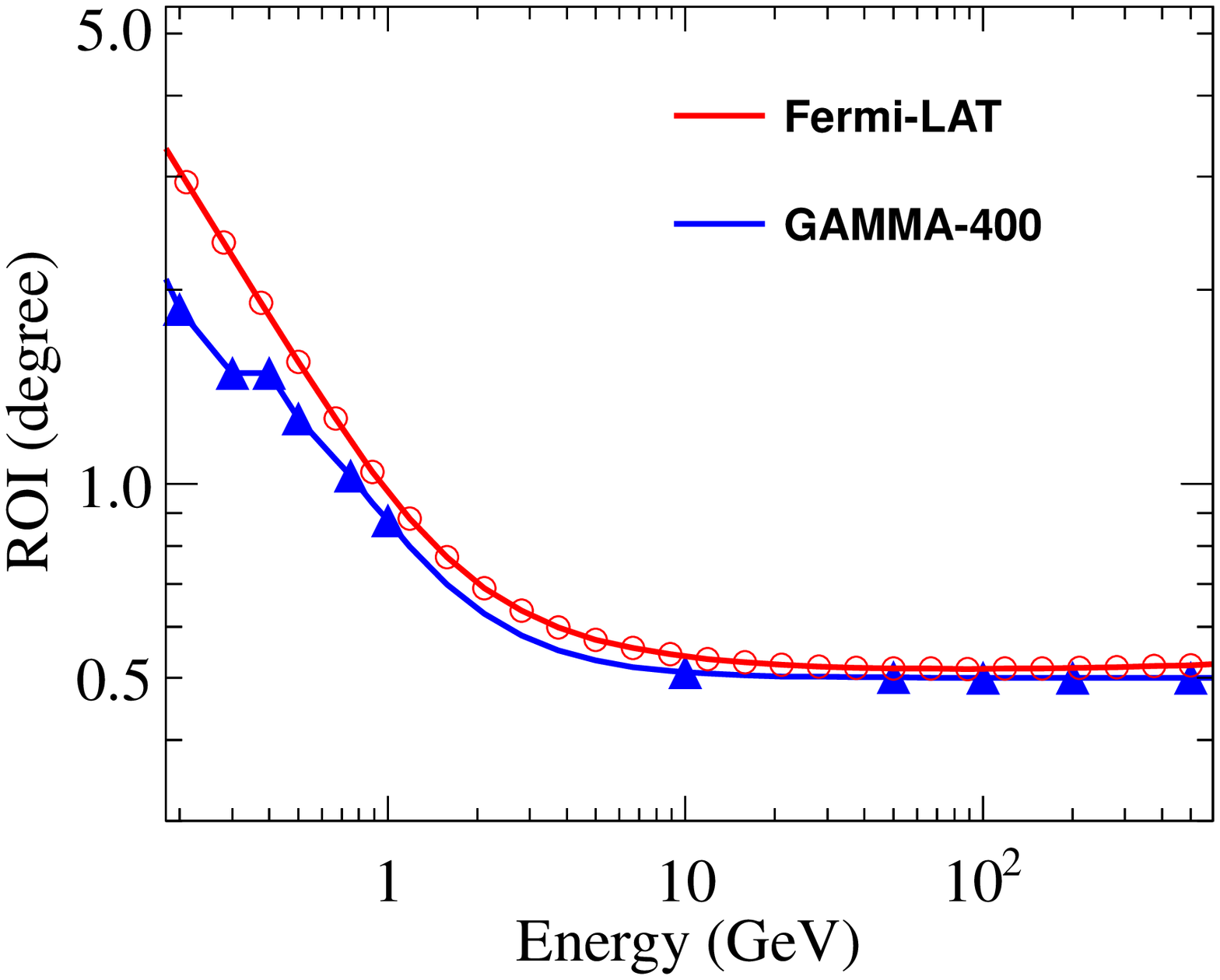}
~
\includegraphics[scale=0.367]{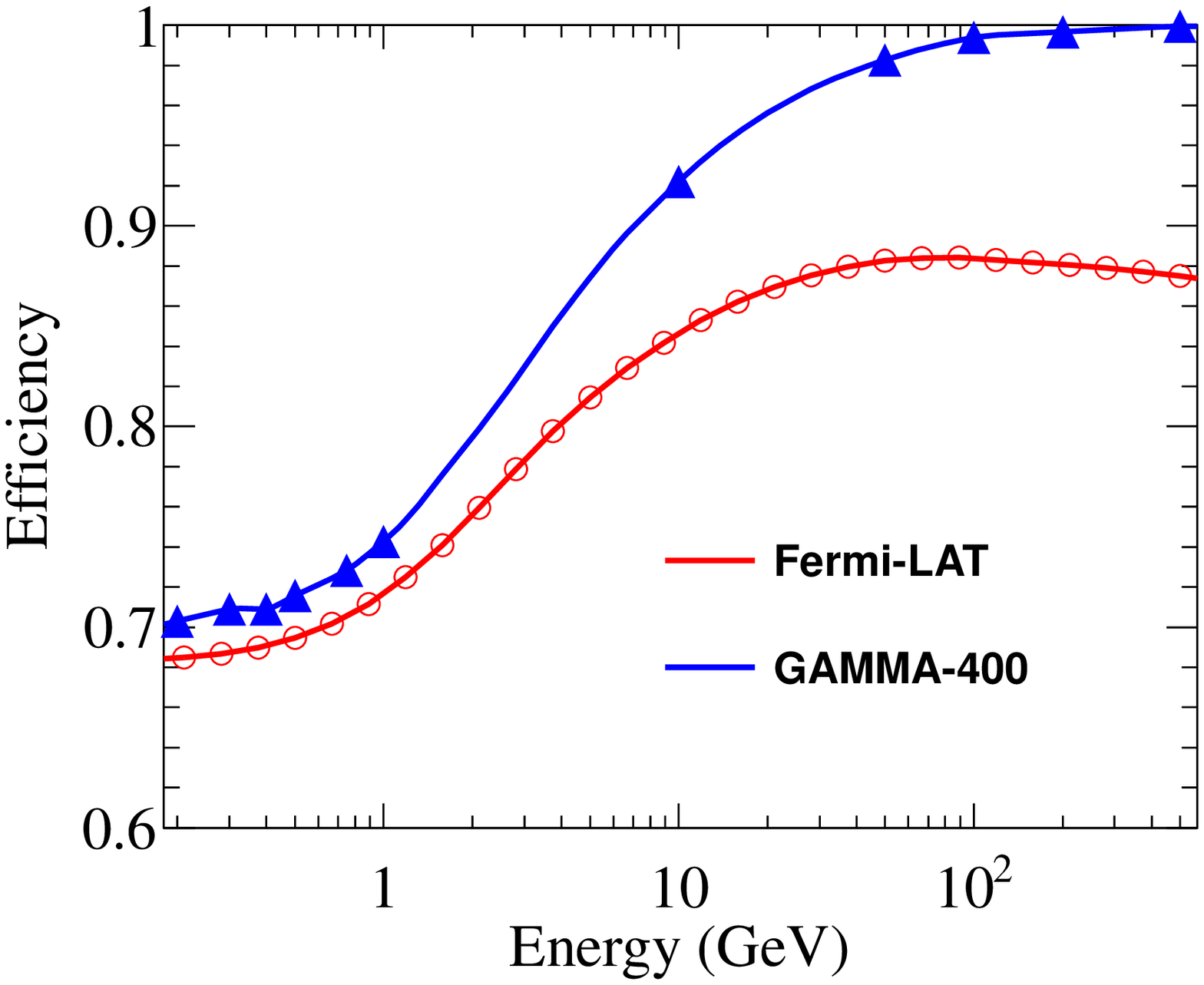}
\caption{\small\sl {\bf Left panel:} The angle defining the region of interest (ROI) used in our analysis as a function of the gamma-ray energy. (The ROI is defined by the circular region with this angle.) {\bf Right panel:} The efficiency factor $\epsilon(\Delta \Omega_i)$ as a function of the energy.}
\label{fig: ROI}
\end{center}
\end{figure}

Using the IRFs presented in Fig.~\ref{fig: IRF}, the number of signal and background events ($S_{ai}$ and $B_{ai}$), which is obtained in actual observation of the dSph `a' at the `i'-th energy bin having the width of $\Delta E_i$, is estimated as
\begin{eqnarray}
S_{ai} &=& t_{\rm obs} \times \int_{\Delta E_i} dE \,
{\cal F}^{{\rm (S)}}_a (E, \Delta \Omega_i) \, A_{\rm eff}(E),
\\
B_{ai} &=& t_{\rm obs} \times \int_{\Delta E_i} dE \,
{\cal F}^{{\rm (B)}}_a (E, \Delta \Omega_i) \, A_{\rm eff}(E),
\end{eqnarray}
where $t_{\rm obs}$ and $A_{\rm eff}(E)$ are the exposure time and the effective area, respectively. We use 25 energy bins with logarithmically equal width in the range of 0.5\,GeV to 500\,GeV, namely the i-th bin has the center vale of $E_i = 0.5 \times 10^{0.125 (i - 1)}$\,GeV with the width of $\Delta E_i/E_i \simeq 0.29$. This choice gives enough large bin-width compared to the energy resolution shown in Fig.~\ref{fig: IRF}. The solid angle $\Delta \Omega_i$ is determined as follows. Though the dark matter profile in each dSph is, as discussed in section~\ref{subsec: A factor}, well concentrated within the circular region with an angular radius of $0.5^{\circ}$, the signal events from the dSph are diffused due to the detector effect. We therefore choose the angle as $\Delta \Omega_i = 2\pi(1 - \cos \alpha_i)$ with $\alpha_i = [(0.5^\circ)^2 + \psi_{68}^2(E_i)]^{1/2}$ to collect most of the signal events. Here, $\psi_{68}$ is the 68\% containment angle (PSF) shown in Fig.~\ref{fig: IRF}. This choice means the region of interest (ROI) is set to be the circular region with the radius of $\alpha_i$. The signal and background fluxes are then given by
\begin{eqnarray}
{\cal F}^{\rm (S)}_a(E, \Delta \Omega_i) =
\epsilon(\Delta \Omega_i) \, \Phi_a(E, \Delta\Omega_i),
\qquad
{\cal F}^{\rm (B)}_a(E, \Delta \Omega_i) =
\Delta \Omega_i \, (d\Phi_a^{\rm B} (E)/d\Omega),
\end{eqnarray}
where $\Phi_a(E, \Delta\Omega_i)$ is the signal flux from the dSph `a' given by the formula~(\ref{eq: flux formula}),\footnote{Since the dark matter profile inside each dSph is concentrated within the circular region with the radius of $0.5^\circ$, the following approximation, $\Phi_a(E, \Delta\Omega_i) \simeq \Phi_a(E, \Delta\Omega_{0.5^\circ})$, is used in our analysis with good accuracy, where $\Delta \Omega_{0.5^\circ}$ denotes the solid angle with the angular radius of $0.5^\circ$.} while $d \Phi_a^{\rm B}(E)/d\Omega$ is the averaged background flux per unit solid angle shown in Fig.~\ref{fig: backgrounds}. The efficiency factor $\epsilon(\Delta \Omega)$ is also introduced in the signal formula to take the loss of highly diffused signal events into account, which is obtained by comparing the signal flux in the ROI with the original one from the dSph. In order to calculate $\epsilon(\Delta \Omega)$, the function provided in reference~\cite{Ackermann:2012kna} (the Gaussian distribution with the width of $\psi_{68}$) was used to describe the angular distribution of the diffusion effect for the Fermi-LAT (GAMMA-400) telescopes. Both the ROI and the efficiency factor are shown in Fig.~\ref{fig: ROI} as a function of the gamma-ray energy. It can be seen that the ROI is governed by the PSF and thus the efficiency factor is about 0.68 when $E < 1$\,GeV, while it is close to one when $E > 10$\,GeV because the PSF becomes negligibly small compared to the angle $0.5^\circ$ (especially for the GAMMA-400).

With the use of signal and background events ($S_{ai}$ and $B_{ai}$) and also the uncertainty of the $J$-factor discussed in section~\ref{subsec: A factor}, the sensitivity to detect the dark matter signal at each telescope, Fermi-LAT and GAMMA-400, can be obtained by the maximum joint likelihood estimation~\cite{Ackermann:2013yva}. The joint likelihood function is constructed by the product of the likelihood function~\cite{Baker:1983tu} for each dSph,
\begin{eqnarray}
{\small
{\cal L}[\langle \sigma v \rangle, \{J_a\}] \equiv \prod_{a,i}
\frac{P\left( N_{ai}; S_{ai}[\langle \sigma v \rangle, J_a] + B_{ai} \right)}
{P\left( N_{ai}; N_{ai} \right)}
G\left( J_a; \log_{10} J^{\rm (obs)}_a, \delta(\log_{10} J^{\rm (obs)}_a) \right),
}
\label{eq: likelihood}
\end{eqnarray}
where $P(N; \lambda)$ and $G(x; \mu, \sigma)$ are the Poisson and the Log Gaussian distributions, respectively, while $\log_{10} J^{\rm (obs)}_a$ and $\delta(\log_{10} J_a^{\rm (obs)})$ are the observed $J$-factor and its error of the dSph `a'. The number of events at the `i'-th energy bin obtained by observing the dSph `a' is denoted by $N_{ai}$. Since we are interested in how severely the annihilation cross section can be constrained with the dark matter mass being fixed in future gamma-ray observations, the number of the signal events $S_{ai}$ depends only on the cross section $\langle \sigma v \rangle$ and the J-factor $J_a$. Because of the same reason, the number of events $N_{ai}$ in our analysis is generated as a mock data following the Poisson distribution with the mean $B_{ai}$. Maximizing the joint likelihood function with respect to the nuisance parameter J, namely $-2\ln {\cal L}(\langle \sigma v \rangle, \{J_{\rm min}\}) + 2\ln {\cal L}(0, \{J_{\rm min}\}) = 2.71$, gives the expected upper limit on the cross section at 95\% confidence level. Here, $\{J_{\rm min}\}$ represents the set of J-factors maximizing the likelihood function for each $\langle \sigma v \rangle$. Using methods developed in this section, we discuss the capability of future dSph observations to explore the wino dark matter in next section.

%% file: Results.tex
\section{Results}
\label{sec: results}

We discuss present and future expected limits on the annihilation cross section of the wino dark matter, utilizing the signal and background gamma-ray fluxes as well as the capability of gamma-ray telescopes mentioned in previous section. The main purpose of discussing the present limit is to confirm whether our method to evaluate detection sensitivity developed in previous section works or not, rather than putting a limit on the cross section utilizing data released by the Fermi-LAT collaboration. Once the method is confirmed to work well, we can robustly estimate how efficiently the wino dark matter can be explored in future by applying the method to expecting data and telescopes. We therefore first summarize the current limit briefly in section~\ref{subsec: present limit}, and evaluate how well our method works in section~\ref{subsec: comparison} by comparing the method with the official one from the Fermi-LAT. After that, we discuss in section~\ref{subsec: future limit} how severely the annihilation cross section can be limited in future, considering both Fermi-LAT and GAMMA-400 telescopes.

\subsection{Present limit}
\label{subsec: present limit}

The most robust limit on the annihilation cross section of the wino dark matter is currently from the four years data of the Fermi-LAT observation~\cite{Ackermann:2013yva}. The limit from each classical dSph (Ursa-Minor, Draco, Sextans, and Sculptor) reported by the Fermi-LAT collaboration is shown in the left panel of Fig.~\ref{fig: present limit}. Here, the dark matter is assumed to annihilate into $W^+ W^-$ with 100\% branching fraction, which is the case we can apply it to the wino dark matter with good precision. Here, it is worth remembering that the limits can be regarded as the robust ones, because the $J$-factors of the classical dSphs are obtained by a robust way, as discussed in section~\ref{subsec: A factor}. It can be seen from the figure that the strongest limit is from the Ursa Minor observation and it gives the limit on the wino dark matter mass as 320\,GeV $\leq m_{\tilde{w}} \leq$ 2.25\,TeV and 2.43\,TeV $\leq m_{\tilde{w}}$ at 95\% confidence level.\footnote{The limit does not significantly altered even if we combine data of all classical dSphs.}

The Fermi-LAT collaboration also provides a more aggressive limit by combining the observational data of fifteen dSphs including eight ultra-faint dSphs, which is also shown in the same panel, and it gives the limit as 390\,GeV $\leq m_{\tilde{w}} \leq$ 2.14\,TeV and 2.53\,TeV $\leq m_{\tilde{w}}$ at 95\% confidence level. It is again worth remembering that the $J$-factors of the ultra-faint dSphs are obtained in a special way, as discussed in section~\ref{subsec: A factor}, and it seems not clear whether or not the limit can be regarded as the robust one. The expected sensitivity from the fifteen dSphs observation, which is also officially given by the Fermi-LAT collaboration, is also shown in the figure as a gray broken-line with the light-green (68\% fluctuation) and the dark-green (95\% fluctuation) bands. The observed limit is slightly deviated from the expected one due to statistical uncertainty (lucky/unlucky factor).

\subsection{Validating our method}
\label{subsec: comparison}

\begin{figure}[t]
\begin{center}
\includegraphics[scale=0.37]{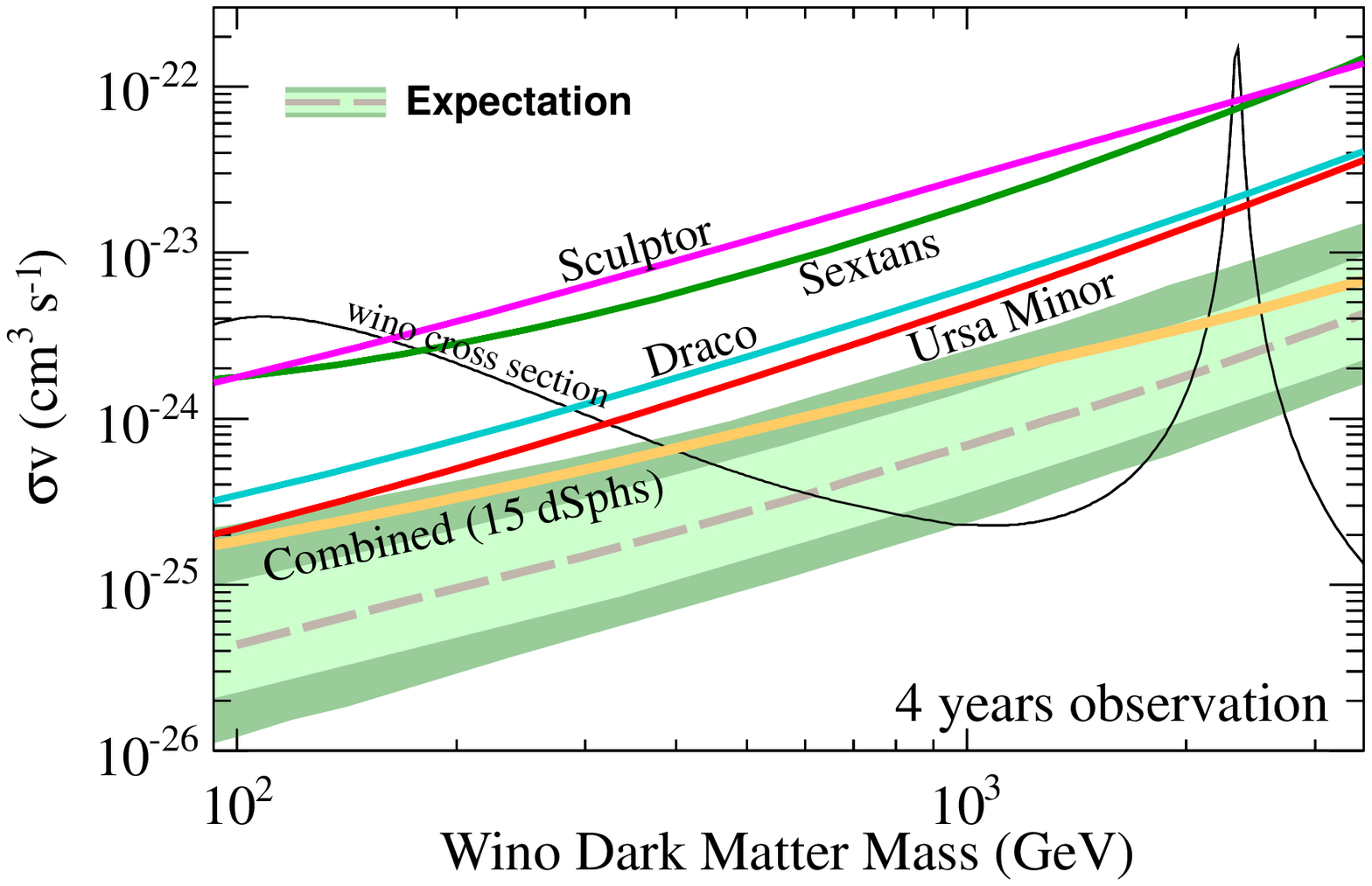}
\includegraphics[scale=0.37]{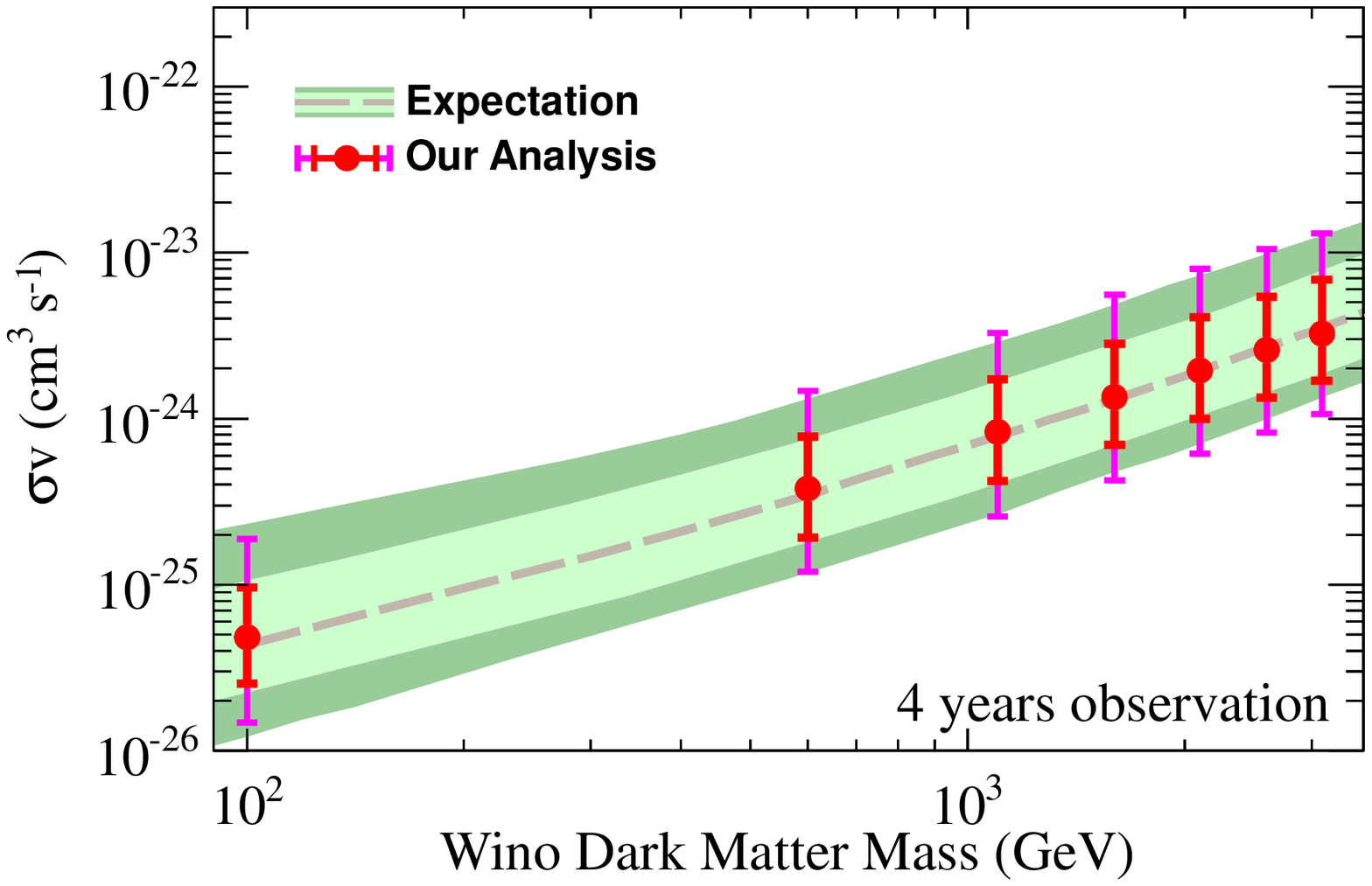}
\caption{\small\sl
{\bf Left panel:} Present limits on $\langle \sigma v \rangle$ from classical dSph observations. An observational limit using the combining data of fifteen dSphs (including eight ultra-faint dSphs) is also shown with the corresponding expected sensitivity given by the Fermi-LAT. {\bf Right panel:} Comparison between our method and Fermi-LAT's one on the expected limit on $\langle \sigma v \rangle$ at 95\% confidence level. See text for more details.}
\label{fig: present limit}
\end{center}
\end{figure}

We are now at the position to discuss how well our method developed in previous section works to give detection sensitivity for the wino dark matter. We consider the eight dSphs discussed in section~\ref{subsec: A factor}. For that purpose, we performed pseudo-experiment $2000$ times, and estimated $N_{ai}$  in equation (\ref{eq: likelihood}) from the obtained (almost Poisson) distribution with the mean value $B_{ai}$. The expectation band (fluctuation) is then obtained by the following procedure, which is also adopted in the Fermi-LAT collaboration: We first calculate the upper limit on the annihilation cross section $\langle \sigma v \rangle$ at 95\% confidence level in each generated mock data with the dark matter mass being fixed. Here, we use the instrumental response functions of the Fermi-LAT assuming data of four years. As a result, we obtain 2000 limits on $\langle \sigma v \rangle$ for each dark matter mass thanks to 2000 generation of mock data. We then calculate the mean value and its 68\% and 95\% fluctuations of the limit by observing the distribution of the 2000 limits.

The median values and their 68\% (95\%) fluctuations of the limit on the cross section $\langle \sigma v \rangle$ are shown in the right panel of Fig.~\ref{fig: present limit} for several dark matter masses, which are depicted as red circles and red (orange) bars, respectively. Those officially from the Fermi-LAT collaboration are also shown in the same panel. It can be seen that not only the median values but also their fluctuations obtained by our method are in good agreement with those from the Fermi-LAT collaboration. It is worth mentioning again that, even if we include the other seven dSphs which are not listed in the tables in the previous section, the result is little changed because the J-factors of these seven dSphs are small compared to the eight dSphs we have used.

\subsection{Expected future limit}
\label{subsec: future limit}

Here, we will give our final results on how widely the mass of the wino dark matter will be explored in (near) future from the gamma-ray observation of dSphs. Two main progresses are expected in this program:\footnote{There may be another progress if we discover new dwarf spheroidal galaxies giving large $J$-factors, for example, by the DES and LSST surveys. See the reference~\cite{He:2013jza} for more details.} One is the accumulation of more data at the Fermi-LAT and the future projected GAMMA-400 telescopes, and another is the improvement of J-factor estimations (especially for ultra-faint dSphs) by obtaining kinematical data of the dSphs accurately. According to these expectations, as near future prospect, we first discuss detection sensitivity (expected future limit) obtained by ten years data of the Fermi-LAT observation assuming $\delta (\log_{10} J) = 0.2$ for ultra-faint dSphs. We next demonstrate detection sensitivity expected from fifteen years data-taking at the Fermi-LAT observation plus ten years data-taking at the GAMMA-400 observation as optimistic but realistic future prospect. Finally, we consider what kind effort and additional observation are required in future to explore entire mass region of the wino dark matter.

\begin{figure}[t]
\begin{center}
\includegraphics[scale=0.5]{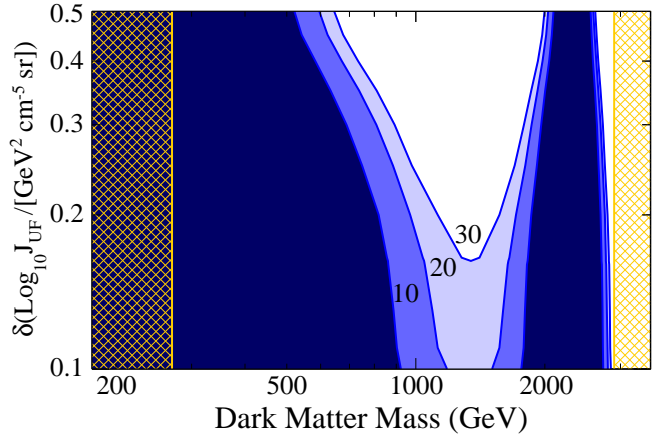}
\caption{\small\sl
Expected limits on the wino mass at 95\% confidence level as a function of $\delta (\log_{10} J_{\rm UF})$ (vertical axis) assuming 10, 20, and 30 years data-taking at the Fermi-LAT. Orange-meshed regions correspond to the limits from the collider search (lower bound) and the thermal relic abundance (upper bound) of the wino dark matter, respectively.}
\label{fig: GvsO}
\end{center}
\end{figure}

\begin{figure}[p]
\begin{center}
\includegraphics[scale=0.67]{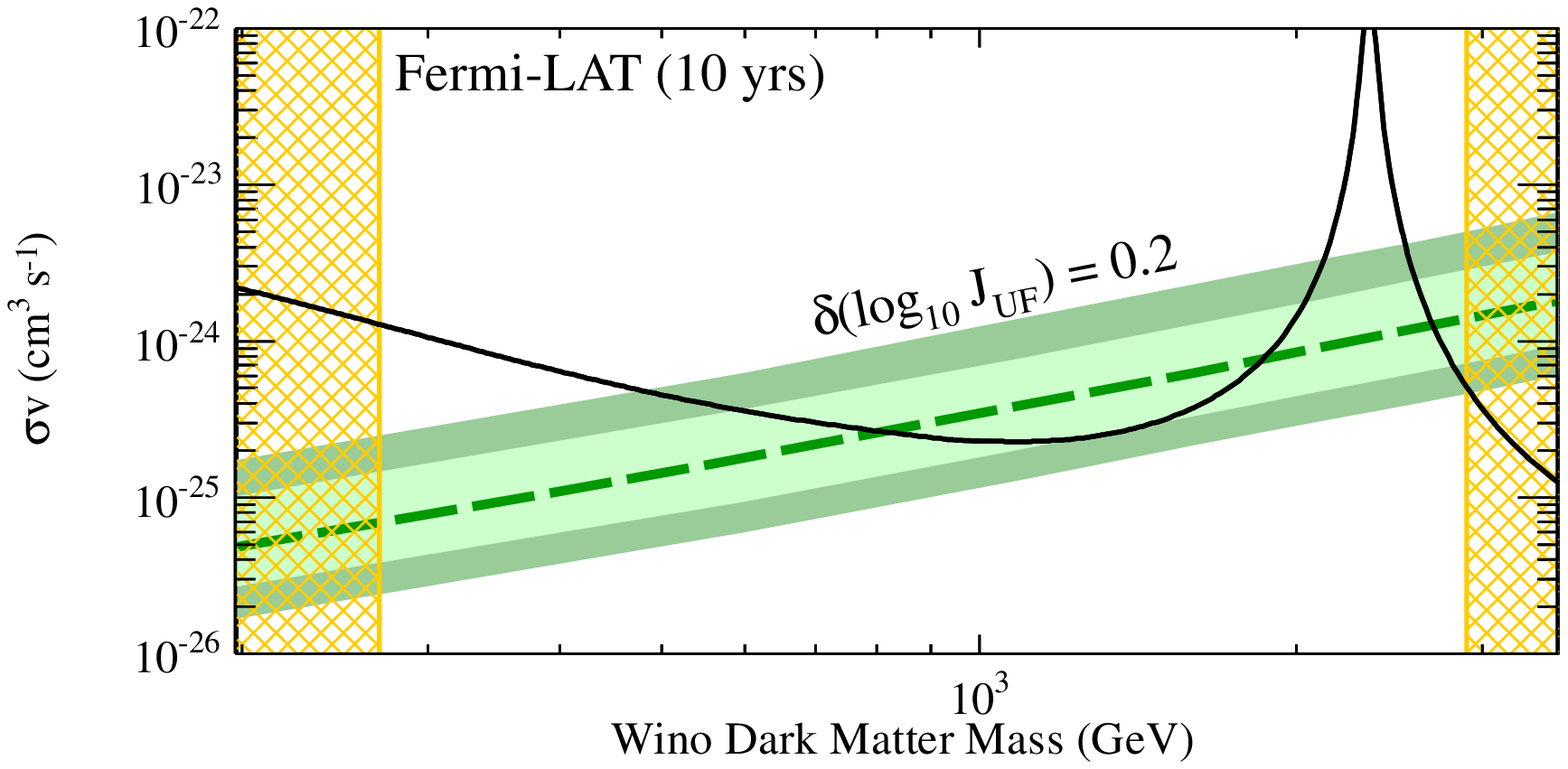}
\includegraphics[scale=0.67]{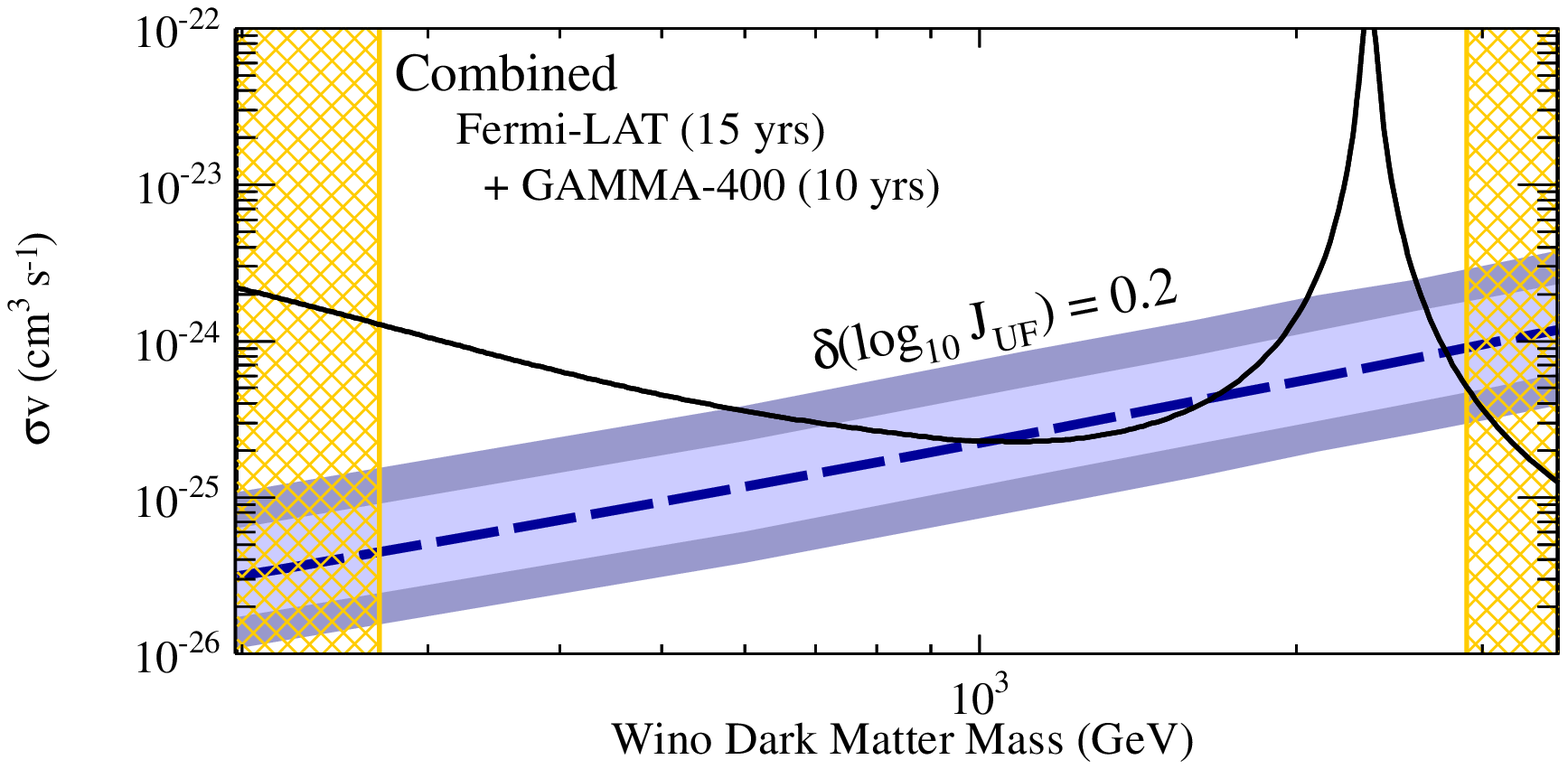}
\includegraphics[scale=0.67]{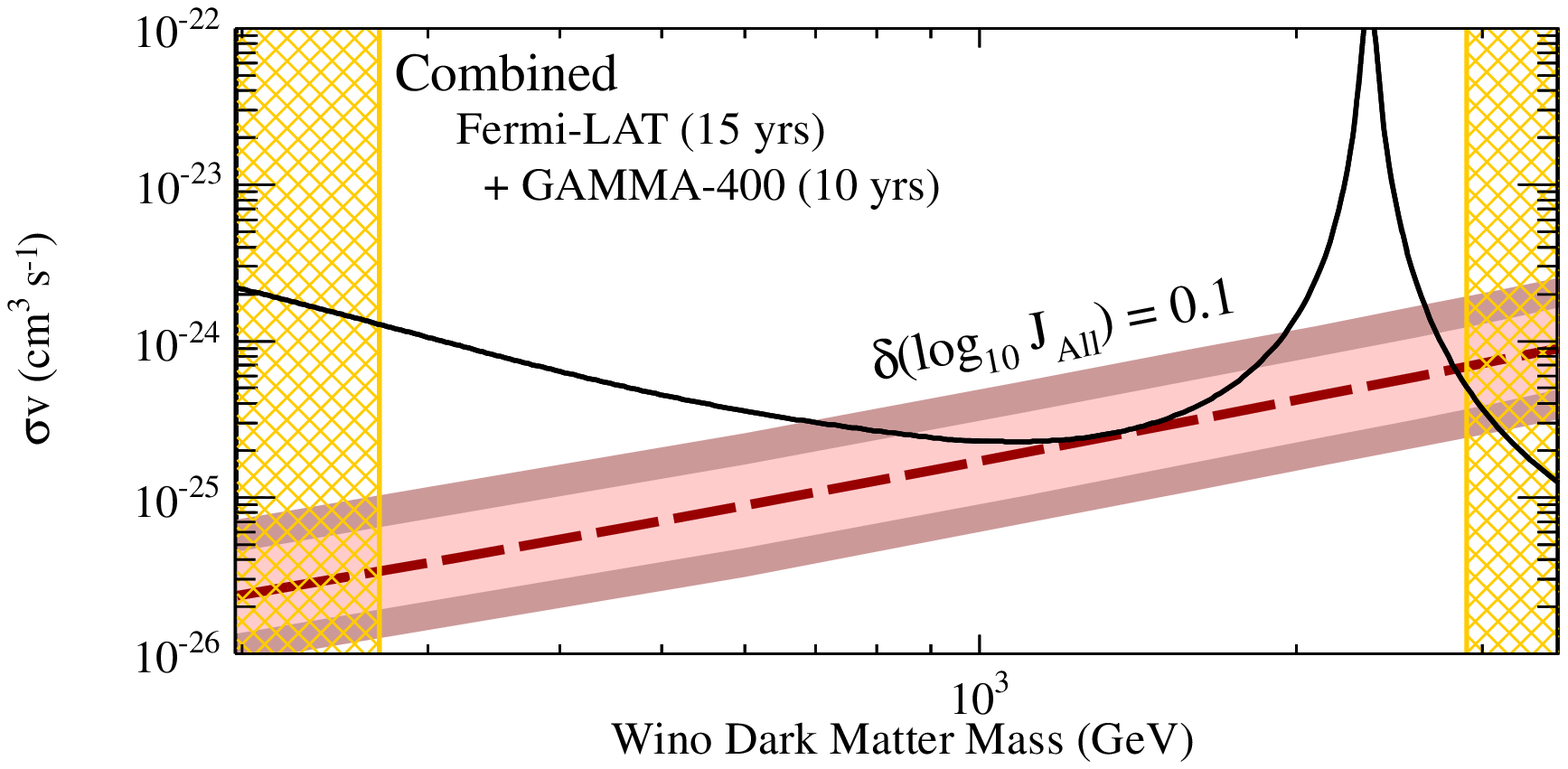}
\caption{\small\sl
Expected future limits on the dark matter annihilation cross section assuming ten years data at the Fermi-LAT and $\delta(\log_{10} J_{\rm UF}) = 0.2$ ({\bf Upper panel}), fifteen years data at the Fermi-LAT plus ten years data at the GAMMA-400 and $\delta(\log_{10} J_{\rm UF}) = 0.2$ ({\bf Middle panel}), and the same as the middle panel but $\delta(\log_{10} J_{\rm All}) = 0.1$ ({\bf Lower panel}). Orange-meshed regions correspond to the limits from the collider search (lower bound) and the thermal relic abundance (upper bound) of the wino dark matter, respectively.}
\label{fig: future limit}
\end{center}
\end{figure}

The expected future limit at the first case is shown in the upper panel of Fig.~\ref{fig: future limit}. Ten years data-taking at the Fermi-LAT observation is the one officially guaranteed by the collaboration~\cite{Baldini:2013gfa}. On the other hand, the logarithmic errors of the J-factors for ultra-faint dSphs, $\delta (\log_{10} J_{\rm UF}) = 0.2$, stems from the fact that current errors for classical dSphs are around 0.2 and from the expectation that deep kinematical survey for ultra-faint dSphs in future will achieve this accuracy. For example, the Prime Focus Spectrograph (PFS) of the SuMIRe Project~\cite{Ellis:2012rn} will be available for this purpose. It is designed to provide a wide field of view ($0.65^\circ$ radius), which is four--five times wider than DEIMOS-KEK~\cite{Faber:2003zz}, keeping an accurate wavelength resolution $R \equiv \lambda/\delta \lambda \sim 3000$. Here, $\lambda$ represents the wavelength of the light covering from 0.38 to 1.3 $\mu$m. Capability of the PFS leads to a large number of kinematical data with high accuracy and the condition $\delta(\log_{10} J_{\rm UF}) = 0.2$ will be satisfied in future. In such a case, from the figure, the wino dark matter with $m_{\tilde{w}} \leq$ 810\,GeV and 1.86\,TeV $\leq m_{\tilde{w}} \leq$ 2.7\,TeV will be explored at 95\% confidence level. It is worth showing that how much of the gains are expected from better measurements of the J-factors in comparison with larger data set from the Fermi-LAT observation. Expected limits on the wino mass at 95\% confidence level are shown in Fig.~\ref{fig: GvsO} as a function of $\delta (\log_{10} J_{\rm UF})$ (vertical axis) assuming 10, 20, and 30 years data-taking. It can be seen from the figure that dramatic improvement of detection sensitivity can be achieved by determining the $J$-factors accurately.

We next consider how the capability of the dSph observation is increased when the GAMMA-400 data becomes available. The expected future limit in this case is shown in the middle panel of Fig.~\ref{fig: future limit}, where fifteen years data at the Fermi-LAT observation plus ten years data at the GAMMA-400 observation is assumed with keeping the errors of the $J$-factors for ultra-faint dSphs being the same as previous case, $\delta(\log_{10} J_{\rm UF}) = 0.2$. The combined analysis of the Fermi-LAT and the GAMMA-400 observations has been performed using the likelihood function constructed by the product of their event likelihoods discussed in previous section. It then turns out from the figure that the wino dark matter with $m_{\tilde{w}} \leq$ 1.0\,TeV and 1.66\,TeV $\leq m_{\tilde{w}} \leq$ 2.77\,TeV will be explored at 95\% confidence level. It is worth mentioning that, though the effective area of the GAMMA-400 telescope is smaller than that of the Fermi-LAT, the accurate point spread function above 10\,GeV guarantees enough efficiency to detect the dark matter. In fact, the capability of the GAMMA-400 observation is almost comparable to that of the Fermi-LAT.

As shown in the middle panel of Fig.~\ref{fig: future limit}, the most of the parameter region for the wino dark matter mass will be covered in future by the dSph observation; it is however not complete and some small regions ($m_{\tilde{w}} \sim$ 1.5\,TeV and 3\,TeV) still remains uncovered. We therefore consider what kind of effort is needed to explore the entire mass region. The simplest solution is, of course, to observe dSphs using telescopes having larger effective area than those of the Fermi-LAT and the GAMMA-400. It is, on the other hand, not obvious whether or not such a costly plan is realized in (near) future. Another solution is to improve estimation of $J$-factors for both classical and ultra-faint dSphs, which requires very precise kinematical data for each dSphs. Such data will be provided if an optical telescope having an wavelength resolution of $R = {\cal O}(10000)$ becomes available. In the bottom panel of Fig.~\ref{fig: future limit}, the expected future limit is shown assuming that the $J$-factors for all the eight dSphs are succeeded to be determined at the level of $\delta (\log_{10} J_{\rm All}) = 0.1$. Here, gamma-ray data is assumed to be the same as previous case (fifteen years data at the Fermi-LAT plus ten years data at the GAMMA-400). It can be seen from the figure that entire mass region (from 270\,GeV to 2.9\,TeV) can be covered in such a case. This fact indicates that not only increasing gamma-ray data but also decreasing the error of the $J$-factor for each dSph are important to cover the entire mass region of the wino dark matter, namely to completely test the high-scale SUSY models.

%% file: Summary.tex
\section{Summary}
\label{sec: summary}

We have thoroughly investigated detection possibility of the wino dark matter in (near) future using the gamma-ray observation of dSphs. Detection or exclusion of the wino dark matter has a strong impact on particle physics, because the high-scale SUSY breaking models, which is now regarded as one of the most promising new physics candidates, predicts the neutral wino as dark matter in most of their parameter region. We have carefully discussed the annihilation of the wino dark matter particle, the dark matter density profile inside each dSph, astrophysical backgrounds against the wino dark matter detection, and the capability of present and future gamma-ray telescopes. All of the issues are mandatory to give robust prospect for the wino dark mater search in (near) future gamma-ray observation.

The mass of the wino dark matter is currently limited as 320\,GeV $\leq m_{\tilde{w}} \leq$ 2.25\,TeV and 2.43\,TeV $\leq m_{\tilde{w}} \leq$ 2.9\,TeV at 95\% confidence level from gamma-ray observation of classical dSphs and the cosmological argument. The limit will be expanded to 810\,GeV $\leq m_{\tilde{w}} \leq$ 1.86\,TeV and 2.7\,TeV $\leq m_{\tilde{w}} \leq$ 2.9\,TeV using ten years data of the Fermi-LAT when the $J$-factors of ultra-faint dSphs are determined with its accuracy of $\delta(\log_{10} J_{\rm UF}) = 0.2$ and no signals are obtained at the observation. When the GAMMA-400 data becomes available, the limit is further improved to 1.0\,TeV $\leq m_{\tilde{w}} \leq$ 1.66\,TeV and 2.77\,TeV $\leq m_{\tilde{w}} \leq$ 2.9\,TeV. Here, fifteen years data of the Fermi-LAT and ten years data of the GAMMA-400 are assumed. In addition, we have considered what kind of effort is eventually needed to search for entire mass region of the wino dark matter. Putting the possibility to have more powerful gamma-ray telescopes aside, the improvement of $J$-factors for both classical and ultra-faint dSphs will play an important role for this purpose. We hope that this result sheds light on the motivation and the clear goal toward kinematical survey of the dSphs with excellent accuracy in (near) future.